# Evidence of Floquet electronic steady states in graphene under continuous-wave mid-infrared irradiation

Yijing Liu[1,#], Christopher Yang[2,#], Gabriel Gaertner[3], John Huckabee[3], Alexey V. Suslov[4], Gil Refael[2], Frederik Nathan[5], Cyprian Lewandowski[4,6], Luis E. F. Foa Torres[7], Iliya Esin[2,8,*], Paola Barbara[1,*], Nikolai G. Kalugin[3,*].

[1] Department of Physics, Georgetown University, Washington, DC 20057, USA.

[2] Department of Physics, IQIM, California Institute of Technology, Pasadena, California 91125, USA.

[3] Department of Materials and Metallurgical Engineering, New Mexico Tech., Socorro, NM 87801, USA.

[4] National High Magnetic Field Laboratory, Tallahassee, Florida 32310, USA

[5] Niels Bohr Institute, University of Copenhagen, Copenhagen 2100, Denmark

[6] Department of Physics, Florida State University, Tallahassee, Florida 32306, USA

[7] Department of Physics, Faculty of Physical and Mathematical Sciences, University of Chile, Santiago 837045, Chile

[8] Department of Physics, Bar-Ilan University, 52900, Ramat Gan, Israel

\# These authors contributed equally

*e-mail: nikolai.kalugin@nmt.edu, paola.barbara@georgetown.edu, iesin@caltech.edu

**Light-induced phenomena in materials can exhibit exotic behavior that extends beyond equilibrium properties, offering new avenues for understanding and controlling electronic phases.[1-3] So far, non-equilibrium phenomena in solids have been predominantly explored using femtosecond laser pulses,[3] which generate transient, ultra-fast dynamics. Here, we investigate the steady non-equilibrium regime in graphene induced by a continuous-wave (CW) mid-infrared laser. Our transport measurements reveal signatures of a long-lived Floquet phase, where a non-equilibrium electronic population is stabilized by the interplay between coherent photoexcitation and incoherent phonon cooling. The observation of non-equilibrium steady states using CW lasers opens a new regime for low-temperature Floquet phenomena, paving the way toward Floquet engineering of steady-state phases of matter.**

Since the inception of quantum theory, light-matter interaction has been a significant source of fascinating discoveries and innovative technologies. In the last few years, Floquet engineering, the use of light to control the properties of a material, emerged as a focus of intense research.[1,2,4-9] This upsurge has been mainly driven by theoretical efforts,[5,6,10-15] with a few key experiments in the condensed matter realm, including the detection of Floquet-Bloch states through time- and angle-resolved photoemission spectroscopy,[16-19] second harmonic generation,[20] as well as the observation of light-induced shifts of exciton resonances in $WS_2$ [21,22] and light-induced Hall effect in graphene.[3] These experiments predominantly explored the transient phenomena induced by ultrafast pulsed lasers, uncovering some of the physics predicted by the Floquet theory. [23] Other studies explored the continuous-wave (CW) regime by inducing discrete Andreev-Floquet bound states in Josephson junctions under CW microwave irradiation.[24,25] However, so far, solid-state experiments in the CW regime have not accessed the Floquet physics of delocalized Bloch states due to challenges related to population transfer effects and heating. In this work, we demonstrate that under intermediate intensity CW mid-infrared (mid-IR) irradiation (see Figure 1 **a**), the electronic population in graphene forms a non-equilibrium steady state with signatures of the underlying single-particle Floquet physics. In particular, the resulting electronic steady state relies on the Rabi-like Floquet gaps and a suppressed density of states (DOS) around electronic energies of half the photon energy $\hbar\Omega$ in graphene's Floquet band structure (see Figure 1 **b-c**). Importantly, we demonstrate that photoinduced transport can be used as a probe of the emergent Floquet steady state in the system. Specifically, we report on signatures of these steady Floquet effects in the gate voltage dependence of the laser-induced longitudinal photoresponses of graphene samples at cryogenic temperatures ranging from 2.5 K to 50 K.

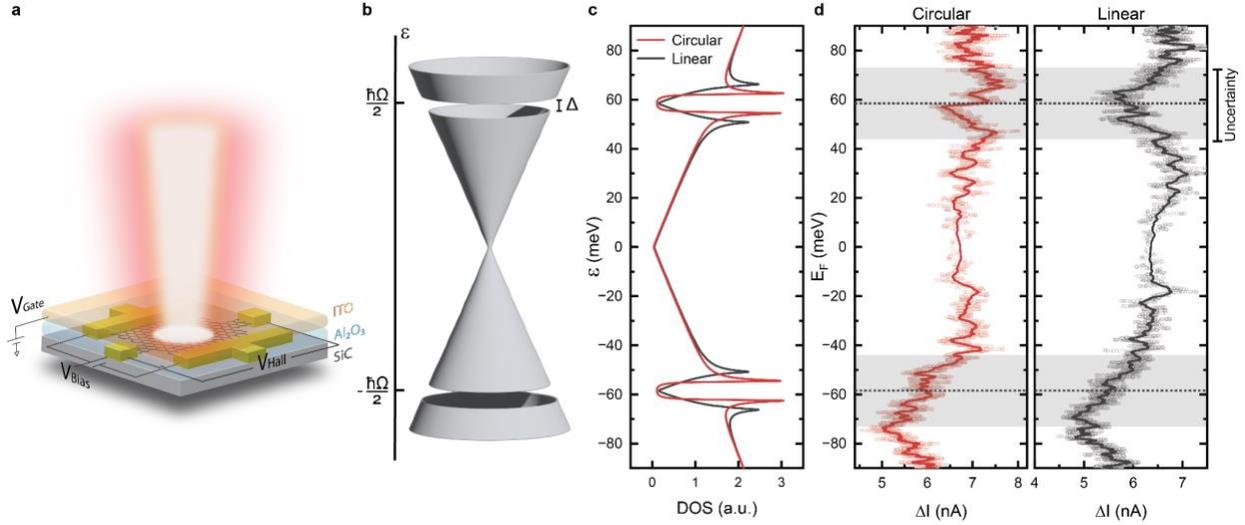

**Figure 1. Device design, Floquet band structure, and longitudinal transport measurements.**
**a** Device layout with two contacts for longitudinal bias and two contacts for measurements of the transverse voltage. The device has a square geometry with a 5-μm side and an indium tin oxide top gate with an Al$_2$O$_3$ dielectric sublayer. **b** Drive-modified Floquet bands of the graphene Dirac cone, which exhibits a Floquet gaps of size Δ at the resonance energies $\varepsilon = \pm\hbar\Omega/2$. **c** Density of states of the Floquet bands calculated numerically for a circularly and linearly polarized laser of power density $P = 3$ mW/μm$^2$. **d** Photo-induced change of source-drain current (*ΔI*) as a function of Fermi energy $E_F$ for circularly and linearly polarized laser irradiation, as measured at a constant source-drain voltage of 6 mV. *ΔI* is measured with a lock-in amplifier using a chopper modulation as a reference. The dotted lines mark the $E_F$ values corresponding to $\pm\hbar\Omega/2$, with the related uncertainty indicated by the gray stripes. The dips in *ΔI* at energies $\pm\hbar\Omega/2$ are much broader than the Floquet gap *Δ* (see panel c) and arise from the non-equilibrium steady-state distribution of electrons. The measurements were performed on sample A at a cryostat temperature of ~3.5K, laser photon energy of 117 meV, and laser power density of 1.1 mW/μm$^2$.

Our graphene devices were fabricated from large-area epitaxial graphene grown on SiC. The typical electron mobility of our material is about 5000 cm$^2$/Vs for the carrier concentration range studied in this work. (Supplementary Figure **S5** shows an image of a cluster with 4 devices.) Each device was a 5 μm × 5 μm graphene square, with two source-drain contacts for the longitudinal electrical transport, two contacts to measure the transverse voltage, and a separate top gate electrode, as depicted in Figure 1 **a**. The top gate comprised a 90-nm-thick dielectric Al$_2$O$_3$ layer and a gate electrode of sputtered 110-nm-thick indium tin oxide (ITO). The properties of our epitaxial graphene, the details of the fabrication process, and the optical properties of the gate electrode are outlined in the Supplementary Information (SI). Our laser operates with a photon energy of $\hbar\Omega = 117$ meV, deliberately chosen to be below the optical phonon generation threshold in graphene (160-200 meV), in contrast to previous experiments that utilized photon energies above this threshold.[3] Importantly, this approach minimizes photo-induced phonon generation, guaranteeing acoustic phonons remain at low cryogenic temperatures and providing cooling channels for photoexcited electrons.

Overall, we measured three top-gated graphene devices on two different chips: samples A and B on the first chip, and sample C on the second chip. The devices were cooled in a closed-cycle optical cryostat with a base temperature of 2.5 K and were irradiated with a 10.6-μm-wavelength CW $CO_2$ laser, providing up to 25 W of linearly polarized light. The laser polarization was controlled with a λ/4 plate. The radiation was delivered using high-power-withstanding molybdenum mirrors, and it was focused on the sample to a spot with a waist of about 50 μm using a ZnSe lens with a focal distance of 50.8 mm. The maximum power used to irradiate the sample was 16 W, yielding a power density $P$ up to 2.6 mW/μm² after considering energy attenuation in the cryostat window and in the top gate of the graphene devices. Further details of the experimental setup are described in the SI.

We performed photoconductive measurements on the Floquet system by measuring the longitudinal current using a lock-in amplifier, with the reference signal to the lock-in provided by a chopper. The source-drain bias was 6mV unless otherwise noted. The chopper modulated the laser beam with a frequency of around 23 Hz and a duty cycle of ~10 %. This procedure provided a direct measurement of the photocurrent, denoted $\Delta I$ and defined as the difference in the source-drain current between the irradiated and equilibrium systems. To explore the doping dependence of the photocurrent, we also extracted the equilibrium Fermi energy, $E_F$, as a function of the applied gate voltage using the classical Hall effect at different gate voltages, as described in detail in the SI.

The experimentally measured photocurrent $\Delta I$ (see Figure 1 **d**) exhibits dips as a function of $E_F$ near doping $E_F \approx \pm 1/2\ \hbar\Omega$ close to the Floquet gaps for both circular and linear polarization of the laser beam. We note that due to the size of the sample chosen to optimize the coupling to the incident beam, [26] the circular polarization might be distorted by the proximity to metallic electrodes[27] (see the SI for details). Nonetheless, the photocurrent dips are robust to changes in polarization and are sensitive only to the electronic photoexcitation rate set by the laser power density and frequency. Importantly, the broadness of the dips, significantly exceeding the width of the Floquet gaps in the graphene density of states (see Figure 1 **d**), cannot be explained with single-particle Floquet physics and instead indicates strong non-equilibrium electronic population effects.

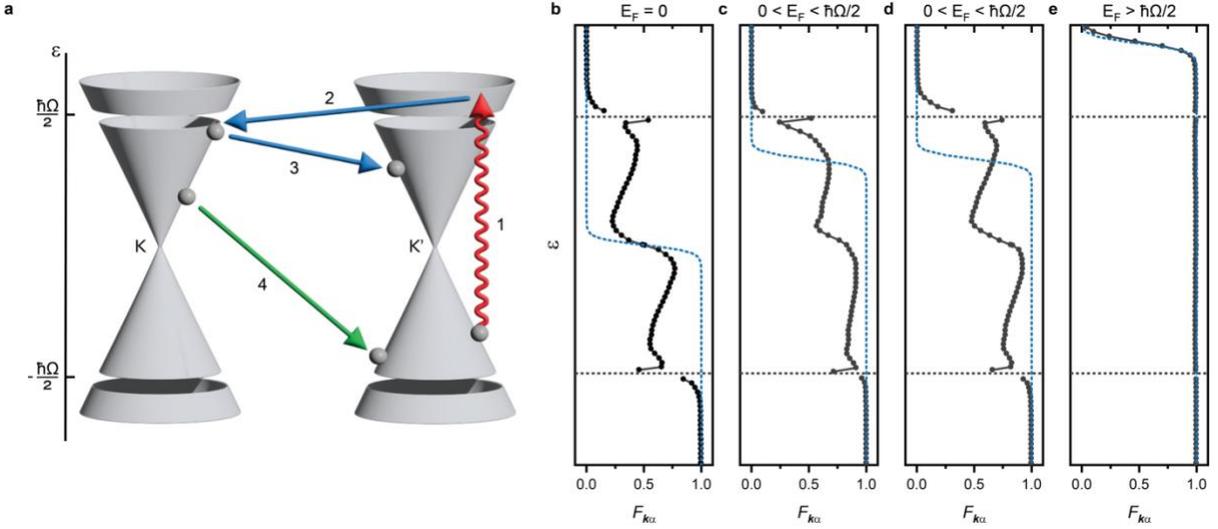

**Figure 2. Expected electronic steady-state distribution. a** Key scattering processes between the graphene $K$ and $K'$ valleys facilitated by photons (arrow 1), surface acoustic phonons (arrows 2,3), and graphene acoustic phonons (arrow 4) that contribute to the steady state. **b** Steady state distribution (black solid curve) for $P = 1.9$ mW/μm² and equilibrium distribution (blue dashed curve) of electrons at doping $E_F = 0$, where the longitudinal photoconductivity is enhanced relative to equilibrium. **c** Steady state distribution (black solid curve) of electrons at doping $E_F = 0.35\ \hbar\Omega$ and $P = 0.8$ mW/μm², where the photoconductivity is suppressed relative to an equilibrium distribution expectation (blue dashed curve). **d** Same as **c** but for a larger power density $P = 1.9$ mW/μm², where the steady state distribution exhibits additional electron density above the Floquet gap. **e** Steady state distribution (black solid curve) of electrons for $P = 1.9$ mW/μm² at doping $E_F = 0.75\ \hbar\Omega$, where the photoconductivity is approximately identical to that of an equilibrium distribution (blue dashed curve), with a slightly raised effective temperature due to multi-photon heating processes.

**Theory of Floquet Steady States**

Under CW illumination, the electronic population forms a non-equilibrium steady state distribution predominantly stabilized by electron-phonon scattering processes (see Figure 2 **a**). To understand the photo-assisted population dynamics, we focus on the low-energy band dispersion $\varepsilon_{k\alpha}$ of graphene, which exhibits two Dirac cones with Fermi velocity $v_F$ corresponding to the graphene $K$ and $K'$ valleys (see Figure 2 **a**). Here, $k$ denotes the crystal momentum, and $\alpha$ enumerates the energy bands. In our theoretical analysis, we focus on circular laser polarization for concreteness. We expect the longitudinal photocurrent to be qualitatively the same for linear polarization parallel to the current. The primary effect of a circularly polarized laser drive is the opening of dynamical gaps at the resonance energies $\pm\hbar\Omega/2$ with a size $\Delta$.[1,4,28] An additional Haldane gap opens at the Dirac point,[1] but the gap is too small to be resolved for the power densities used in the experiment, and is therefore not shown in Figure 2 **a**. The dominant photoexcitation process corresponds to an excitation of the electron to a virtual state (arrow 1) through photon absorption followed by the phonon-emission process relaxing the electron to the conduction band (arrow 2). The electronic population excited by this process is subsequently spread across a small window of energies assisted by the emission of low-energy acoustic phonons (arrow 3) and can be subsequently relaxed through additional acoustic phonon emission processes (arrow 4) back into the valence band. Ultimately, in the steady state, the excited electronic occupation in the conduction band is set by the balance between phonon-assisted photo-excitations (arrows 1-3) and phonon-assisted cooling processes (arrow 4). In Figure 1 **c** and throughout our theoretical calculations, we include two phonon branches: a slow surface acoustic phonon branch with Rayleigh wave speed 1.3 km/s (arrows 2 and 3 in Figure 1 **c**) and a faster graphene acoustic phonon branch with speed 11 km/s (arrows 4 in Figure **c**).[29] The surface acoustic phonon speed used in our simulations captures the order of magnitude of the Rayleigh wave speed of the SiC.[30] To find the steady state electronic population $F_{k\alpha}$, we solved numerically the full Floquet-Boltzmann equation, which includes all the microscopic electron-phonon scatterings between Floquet-Bloch states in the laser-driven graphene (see SI for details).[31-34] The resulting steady state occupations at various electronic dopings are plotted in Figures 2 **b-e**, and the photocurrent is calculated from the steady state occupations using linear response theory in the weak source-drain bias regime. In our theory, we also account for charge puddles and a small, intensity-dependent lattice temperature difference between the driven and undriven systems. We estimate the conductivity in the presence of charge puddles using a phenomenological formula detailed in the SI.

Remarkably, the steady state distributions denoted by black curves in Figures 2 **b-e** exhibit multiple step-like features associated with large $|\partial F_{k\alpha}/\partial \varepsilon_{k\alpha}|$, resembling effective Fermi surfaces, which significantly affect transport. The electron-like effective Fermi surfaces ($\partial F_{k\alpha}/\partial \varepsilon_{k\alpha} < 0$) contribute to a longitudinal current parallel to the direction of an applied electric field, while hole-like effective Fermi surfaces ($\partial F_{k\alpha}/\partial \varepsilon_{k\alpha} > 0$) suppress the overall photocurrent by contributing current antiparallel to the applied field. This is in strict contrast to an equilibrium distribution, where a single Fermi surface appears in the electronic distribution (see blue dashed curves in Figures 2 **b-e**), and only electrons near the Fermi surface contribute to the longitudinal transport. The Fermi surfaces in the distribution predominantly arise from the scattering processes 1-4 sketched in Figure 2 **a**. Most notable is the non-equilibrium distribution for an electron-doped system $0 < E_F < \hbar\Omega/2$ (see Figure 2 **c-d**), where the equilibrium Fermi surface located at finite density of states (see blue dashed curve) is separated into three electron-like Fermi surfaces (see black solid curve) which are located near the dips of the density of states and therefore contribute very few electronic carriers. In particular, two electron-like Fermi surfaces are centered at the Floquet gaps, where electronic group velocities and density of states are reduced, and another is positioned at zero energy, where the density of states vanishes. The reduced concentration of electronic carriers gives rise to a suppressed conductivity in the driven system. This phenomenon persists for both weak and strong driving amplitudes, as shown in Figures 2 **c** and **d**, respectively. For doping above resonance $E_F > \hbar\Omega/2$, processes 1-4 are Pauli-blocked, giving rise to an equilibrium-like distribution (see Figure 2 **e**).

**Optically-Controlled Photoresponse**

Having understood the interplay of the Floquet bands, steady state occupation, and photocurrent $\Delta I$, we now compare the theoretically calculated and experimentally measured value of $\Delta I$ and discuss the tunability of the photoresponse by the laser power. In Figure 3 **a**, we focus on the experimentally measured photocurrent dip at negative $E_F$ and plot the photocurrent for various laser power densities. In Figure 3 **b**, we show the theoretically calculated photocurrent for the same power densities. Let us discuss a few salient features captured by the theoretical model. A key feature is the broad width of the photocurrent dip, which we estimate by fitting the photocurrent to a Gaussian-like function detailed in the Methods section. The amplitude of the Gaussian $a_0$ quantifies the depth of the photocurrent dip, while the full width at half maximum (FWHM) quantifies its width. Notably, our theoretical prediction of the decrease in $a_0$ with the power density $P$, agrees with the experimental observation (see Figure 3 **c**), indicating a power dependence of the steady state distribution. The reduction of $a_0$ with $P$, arises due to the increase of the photoexcited carrier density above the Floquet gap, which scales as $\sim P^{1/2}$, effectively increasing the photocurrent (cf. the shift of the effective Fermi surface near $\hbar\Omega/2$ in Figures 2 **c** and 2 **d**).[28] Finally, we note that the FWHM of the photocurrent dip observed in the experiment and predicted by the theory are both broader than the size of the Floquet gap, see Figure 3 **d**, indicating a Floquet-induced electronic population inversion in the steady state. Finally, we note that $\Delta I > 0$ for $0 < E_F < \hbar\Omega/2$ due to charge puddle and lattice heating effects (see the SI for more details).

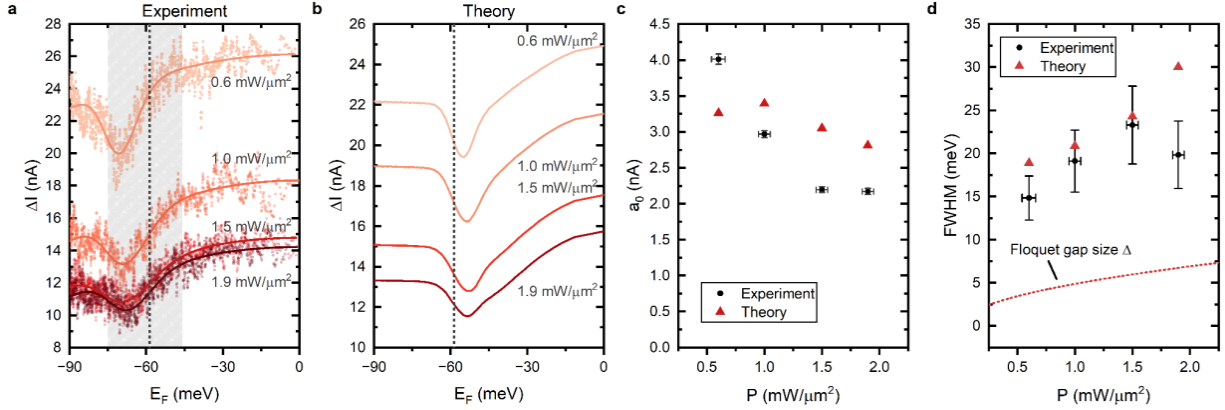

**Figure 3. Transport signatures. a** Photo-induced change of source-drain current ($\Delta I$) as a function of $E_F$ under irradiation at various peak power densities, as measured on sample C at a constant source-drain voltage of 6 mV and a cryostat temperature of 3.2 K. The photon energy is $\hbar\Omega$ =117 meV. The laser beam is circularly polarized. **b** Theoretically predicted $\Delta I$ as calculated from the Floquet Boltzmann equation. **c** Depth $a_0$ of the photocurrent dip, as calculated from a Gaussian-like fit (see Methods section). For both the theoretical and experimental data, $a_0$ decreases with the power density $P$ for large $P$ due to enhanced heating processes in the Floquet steady state. **d** Experimental and theoretical FWHM of the photocurrent dip and the Floquet gap size, $\Delta$, as a function of the power density. The FWHM exceeds $\Delta$, indicating the emergence of photoexcited electrons in the non-equilibrium Floquet steady state.

To further verify that the transport signatures observed in the experiment arise from photo-induced dynamics, we explore the experimentally measured photo-response for out-of-focus lasers, which reduce the irradiation power density by roughly a factor of three. Figure 4 **a** demonstrates that the visibility of the photocurrent dip is significantly suppressed as the power density is reduced. This behavior, reproduced in theory (see Figure 4 **b**), reflects the suppressed probability of drive-induced photoexcitation processes under weak laser irradiation.

Next, we explore the lattice temperature dependence of the electronic steady state and conductivity. Figures 4 **c** and **d**, respectively, show the measured and the theoretically predicted $\Delta I$ as a function of $E_F$, for several values of the lattice temperature. At higher temperatures, the distribution of the photoexcited electrons spreads over larger energy support, relaxing the sharp energy-momentum bottlenecks in the steady state distribution (see processes 1-4 in Figure 2 **a**).

As a result, the dips in *ΔI* become less pronounced, virtually disappearing around 20 K. The discrepancy of the temperature dependence around charge neutrality between the experimental and theoretical plots is attributed to the finite-temperature physics of the charge puddles, which could modify the relaxation time of electrons at different temperatures under weak source-drain bias,[35] which is not captured in the theory. The strong lattice temperature dependence of both the theoretically calculated and experimentally observed photocurrent, however, highlights the role of low phonon temperatures in stabilizing low-temperature electronic phenomena of Floquet effects in our system.

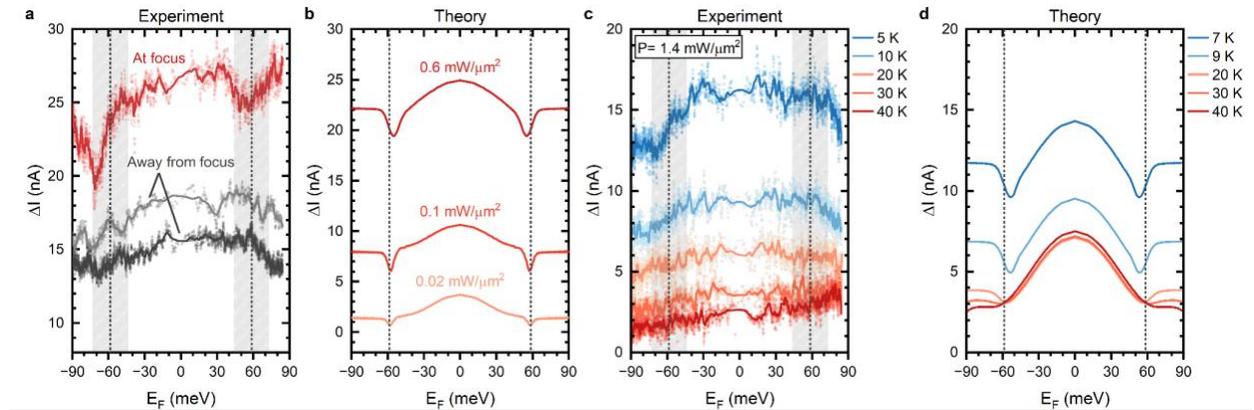

**Figure 4. Laser intensity and temperature dependence of the longitudinal conductance.**
**a** Longitudinal photocurrent as a function of $E_F$ at a fixed source-drain voltage of 6 mV, with laser spot in focus and defocused. For all three curves, the laser power is ~3.6 W, corresponding to a peak power density of 0.6 mW/μm² at focus. Under defocused irradiation, with the laser spot slightly moved to the side, the power density drops by ~3 times (gray curves). Circular polarization of 10.6 mm wavelength laser radiation. The cryostat temperature stabilized at ~3.4 K **b** Theoretically calculated photocurrent for weak drives, exhibiting shallower dips near resonance as a function of decreasing driving power, in agreement with panel **a**. **c** Temperature dependence of the longitudinal photocurrent as a function of $E_F$. The photocurrent decreases, and the dips at $E_F = \pm 1/2\ \hbar\Omega$ disappear at high temperatures. The laser beam is circularly polarized. **d** Theoretically calculated photocurrent for laser power density 1.4 mW/μm², the photocurrent dip becomes less visible at higher temperatures.

We note that the photoconductivity dip has been reproduced for other samples, which we present in the SI. Different samples with different charge inhomogeneities showed similar widths of the photocurrent dip. This observation is consistent with the emergence of a Floquet steady state, which is predicted to be weakly sensitive to disorder. The positions of the dips in the electron and hole regions of the gate dependence are also slightly asymmetric, similar to particle-hole asymmetric results reported in McIver et al.[3]

Finally, we discuss the transverse photoconductivity of the system. Our Hall voltage measurements displayed a laser helicity-dependent component on the $\mu V$ scale, which amounts to a Hall conductivity of the order of $\sigma_{xy}^{Hall} \sim 10^{-3} e^2/h$, see extended data Figure **E1**. This range of values undershoots the prediction of $\sigma_{xy}^{Hall} \sim e^2/h$, from our theoretical model, which is mainly designed as a simple model for the longitudinal component of the photocurrent. We believe that a key reason for this discrepancy is the distortion of the circular polarization by the metallic contacts, leading to regions of the sample with nearly linear polarization[27] (see SI for details). In these regions, the Berry curvature can have large off-diagonal elements, reducing the photoresponse. Additionally, our theoretical calculation of the Hall voltage does not account for electron-electron interactions and charge puddles. Electron-electron scattering could suppress the Hall conductivity by raising the effective temperature of the steady-state electron-like Fermi surfaces (see Figure 2 **b-e**) near the resonant Floquet gaps.[32,34] Charge puddles could further suppress the Hall voltage by producing regions in the sample with low conductivity, which exhibit weak longitudinal source drain current and heavily suppressed local Hall voltage. More details of the transverse photoresponse measurements are provided in the extended data and in the SI.

**Conclusion**

In this work, we have explored the transport properties of graphene driven by a CW mid-infrared laser. In such metallic systems, the Floquet dressing of single-particle bands by resonant periodic drives is intertwined with many-body effects, such as electron scattering and population dynamics. Notably, we demonstrate that the interplay between these processes at intermediate driving intensities and cryogenic temperatures facilitates the formation of low-entropy electronic steady states. These steady states emerge from a cascade of photo-assisted electron-phonon scattering events that effectively cool the photoexcited electrons to low entropy states.

Our findings suggest that, in graphene on SiC, these processes are dominated by the emission of surface acoustic phonons at the graphene-SiC interface, along with acoustic phonons in the graphene itself. Furthermore, the formation of these steady states is crucially dependent on the underlying single-particle Floquet physics, such as the presence of Floquet gaps in the single-particle spectrum and scattering into replica Floquet bands. The presence of the Floquet steady states is evidenced in our transport measurements, which exhibit a pronounced dip in longitudinal photoconductivity. The characteristics of this dip—its position, depth, and width as functions of doping—align closely with predictions from Floquet many-body theory. Our results underscore the potential of Floquet engineering in metallic systems for sustained operation, paving the way for the creation of novel Floquet steady-state phases, such as drive-induced symmetry broken phases,[36] laser-induced flat bands,[34,37,38] and optically-controlled topological transport.[33,34]

**Materials and Methods**

Experiment

The CW source was a 10.6μm-wavelength, air-cooled Synrad J48-2 $CO_2$ laser, providing up to 25W of linearly polarized light, modulated with a chopper (Scitec Instruments 300CD) with 27 Hz modulation frequency. The circular polarization of the laser radiation was controlled by a custom-made zero-order λ/4 plate, provided by Optogama UAB. For the delivery of the optical beam, we used molybdenum mid-IR high power mirrors.

The samples were cooled down to 2.5 K with a closed-cycle cryostat Oxford Instruments Optistat AC-V14. For optical access to our samples, we used a 2-mm thick ZnSe window. The beam delivery was done using an Edmund Optics cage focusing system mounted on a custom-made mechanical attachment to the cryostat. For beam focusing, we used a 50.8-mm focal distance, 1-inch diameter, ZnSe lens. Focusing and alignment of the mid-IR beam was controlled by a system of micrometers.[39] The estimation of the beam diameter at the lens focus is based on a Gaussian beam with our experimental parameters (18-19 mm laser beam waist before lens and 50.8 mm lens focal distance), and it yields ~35-38 μm. Experimentally, we were able to achieve focusing down to a 50 μm spot. This was confirmed during the fine alignment of the laser beam: the magnitude of *ΔI* was changing from the noise level to a maximum signal and then back to the noise level (about 10-12-times drop of signal magnitude) within 4-5 readout thimble divisions of in-plane alignment micrometers (e.g., within 40-50 μm distance). The maximum applied laser power in our experiments was 16 W, corresponding to a peak power density ~ 8 mW/μm$^2$ for ~50 μm-diameter beam spot. Taking into account the transmission coefficients of the cryostat window and the gate electrode, we estimate the applied laser intensity as ~ 2.6 mW/μm$^2$.

The measurements of photoinduced currents and voltages were performed using a home-made bias box, an HP 6177C DC current/voltage source (for generation of gate voltages), a National Instruments BNC-2110 junction box, an Ametek 5110 lock-in amplifier, a DL Instruments 1211 current preamplifier, and a custom-made differential voltage preamplifier. The data collection system was controlled using a customized LabVIEW-based program.

The calibration of gate efficiencies of our samples (calibration of gate voltage in units of electron Fermi energy) was performed using a 9 Tesla Quantum Design Physical Property Measurement System at the NHMFL in Tallahassee. The details about gate efficiency calibration are provided in the SI.

Device fabrication

For the device fabrication we adapted the process developed by Yang et al.[40] to electron-beam lithography (EBL), as described in Ref.[41] with additional lithography steps to deposit and pattern a top gate. The epitaxial graphene on SiC was purchased from Graphene Waves. The details of the device fabrication are provided in the SI.

Fitting

We estimate the full width at half maximum (FWHM) of the photocurrent dip by fitting the photocurrent to the function $\Delta I_{\text{fit}} = a_0 \exp\left[-(V_g - B)^2/2\sigma^2\right] + CV_g + D$, where $a_0, B, \sigma, C$, and $D$ are fitting parameters, and $V_g$ is the gate voltage. We extract the full width half maximum using the relation $\text{FWHM} = E_F(B + \sqrt{2 \ln 2}\,\sigma) - E_F(B - \sqrt{2 \ln 2}\,\sigma)$, where $E_F(V_g) = \hbar v_F \sqrt{\pi k (V_g - V_d)}$. Here, $k$ is estimated from Hall measurements, and $V_d$ is estimated to be the gate voltage at which $\sigma_{xx}$ is minimized (see the SI).

**Acknowledgments**

We acknowledge support from NSF (projects DMR CMP #2104755, DMR CMP #2104770, and OSI #2329006), FondeCyT (Chile) through grant number 1211038, and the Institute for Quantum Information and Matter, an NSF Physics Frontiers Center (PHY-2317110). C.Y. gratefully acknowledges support from the DOE NNSA Stewardship Science Graduate Fellowship program, which is provided under cooperative Agreement No. DE-NA0003960. G.R. and I.E. are grateful to the AFOSR MURI program, under agreement number FA9550-22-1-0339, as well as the Simons Foundation. Part of this work was done at the Aspen Center for Physics, which is supported by the NSF grant PHY-1607611. F.N. gratefully acknowledges support from the Carlsberg Foundation, grant CF22-0727. C.L. was supported by start-up funds from Florida State University and the National High Magnetic Field Laboratory. The National High Magnetic Field Laboratory (NHMFL) is supported by the National Science Foundation through NSF/DMR-1644779, NSF/DMR-2128556 and the State of Florida. L. E. F. F. T. acknowledges partial support from the EU Horizon 2020 research and innovation program under the Marie-Sklodowska-Curie Grant Agreement No. 873028 (HYDROTRONICS Project), and of The Abdus Salam International Centre for Theoretical Physics and the Simons Foundation.

The authors thank Dr. Michael Jackson, Dr. Yanfei Yang, Eli Adler, DaVonne Henry, Amjad Alqahtani, and Thy Le for helpful discussions, Taylor Terrones, Michael Chavez, and Leon Der for help with our experimental setup, and Dr. David Graf for help with experiments at NHMFL.

**Author Contributions**

N.G.K., L.E.F.F.T., and P.B. planned the project. Device fabrication was performed by Y.L. Optical systems were designed by N.G.K., A.V.S., G.G., J.H., P.B., Y.L., and were assembled and tested by N.G.K, A.V.S., G.G., and J.H. with the help of machine shops of NMT, GU, and NHMFL. Measurements were made by Y.L., P.B., A.V.S., and N.G.K. L.E.F.F.T. carried out theoretical calculations of the Floquet density of states. C.Y. carried out the steady-state Floquet simulations supervised by G.R., F.N., C.L., and I.E. All authors contributed to the discussion of the results and the manuscript preparation.

**Additional Information**

The authors declare no competing financial interests.

**Data available with the paper or supplementary information**
The authors declare that the data supporting the findings of this study are available within the paper and its SI. Data sets generated during the current study are available from corresponding authors on reasonable request.

**Code availability**
Computer code generated during the current study is available from corresponding authors on reasonable request.

# References

1　Oka, T. & Aoki, H. Photovoltaic Hall effect in graphene (vol 79, 081406, 2009). *Physical Review B* **79** (2009). https://doi.org/10.1103/PhysRevB.79.169901
2　Lindner, N. H., Refael, G. & Galitski, V. Floquet topological insulator in semiconductor quantum wells. *Nature Physics* **7**, 490-495 (2011). https://doi.org/10.1038/nphys1926
3　McIver, J. W. *et al.* Light-induced anomalous Hall effect in graphene. *Nature Physics* **16**, 38-+ (2020). https://doi.org/10.1038/s41567-019-0698-y
4　Kitagawa, T., Oka, T., Brataas, A., Fu, L. & Demler, E. Transport properties of nonequilibrium systems under the application of light: Photoinduced quantum Hall insulators without Landau levels. *Physical Review B* **84** (2011). https://doi.org/10.1103/PhysRevB.84.235108
5　Oka, T. & Kitamura, S. Floquet Engineering of Quantum Materials. *Annual Review of Condensed Matter Physics, Vol 10* **10**, 387-408 (2019). https://doi.org/10.1146/annurev-conmatphys-031218-013423
6　Rudner, M. S. & Lindner, N. H. Band structure engineering and non-equilibrium dynamics in Floquet topological insulators. *Nature Reviews Physics* **2**, 229-244 (2020). https://doi.org/10.1038/s42254-020-0170-z

7	Perez-Piskunow, P. M., Usaj, G., Balseiro, C. A. & Torres, L. Floquet chiral edge states in graphene. *Physical Review B* **89** (2014). https://doi.org/10.1103/PhysRevB.89.121401

8	Jotzu, G. *et al.* Experimental realization of the topological Haldane model with ultracold fermions. *Nature* **515**, 237-U191 (2014). https://doi.org/10.1038/nature13915

9	Seetharam, K., Titum, P., Kolodrubetz, M. & Refael, G. Absence of thermalization in finite isolated interacting Floquet systems. *Physical Review B* **97** (2018). https://doi.org/10.1103/PhysRevB.97.014311

10	Kundu, A., Fertig, H. A. & Seradjeh, B. Effective Theory of Floquet Topological Transitions. *Physical Review Letters* **113** (2014). https://doi.org/10.1103/PhysRevLett.113.236803

11	Torres, L., Perez-Piskunow, P. M., Balseiro, C. A. & Usaj, G. Multiterminal Conductance of a Floquet Topological Insulator. *Physical Review Letters* **113** (2014). https://doi.org/10.1103/PhysRevLett.113.266801

12	Dehghani, H., Oka, T. & Mitra, A. Out-of-equilibrium electrons and the Hall conductance of a Floquet topological insulator. *Physical Review B* **91** (2015). https://doi.org/10.1103/PhysRevB.91.155422

13	Rudner, M. S. & Song, J. C. W. Self-induced Berry flux and spontaneous non-equilibrium magnetism. *Nature Physics* **15**, 1017-+ (2019). https://doi.org/10.1038/s41567-019-0578-5

14	Bajpai, U., Ku, M. J. H. & Nikolic, B. K. Robustness of quantized transport through edge states of finite length: Imaging current density in Floquet topological versus quantum spin and anomalous Hall insulators. *Physical Review Research* **2** (2020). https://doi.org/10.1103/PhysRevResearch.2.033438

15	Sato, S. A. *et al.* Microscopic theory for the light-induced anomalous Hall effect in graphene. *Physical Review B* **99** (2019). https://doi.org/10.1103/PhysRevB.99.214302

16	Wang, Y. H., Steinberg, H., Jarillo-Herrero, P. & Gedik, N. Observation of Floquet-Bloch States on the Surface of a Topological Insulator. *Science* **342**, 453-457 (2013). https://doi.org/10.1126/science.1239834

17	Zhou, S. H. *et al.* Pseudospin-selective Floquet band engineering in black phosphorus. *Nature* **614**, 75-+ (2023). https://doi.org/10.1038/s41586-022-05610-3

18	Mahmood, F. *et al.* Selective scattering between Floquet-Bloch and Volkov states in a topological insulator. *Nature Physics* **12**, 306-U137 (2016). https://doi.org/10.1038/nphys3609

19	Merboldt, M. *et al.*   (arXiv:2404.12791v1   [cond-mat.mes-hall], 2024).

20	Shan, J. *et al.* Giant modulation of optical nonlinearity by Floquet engineering. *NATURE* **600**, 235-+ (2021). https://doi.org/10.1038/s41586-021-04051-8

21	Kobayashi, Y. *et al.* Floquet engineering of strongly driven excitons in monolayer tungsten disulfide. *Nature Physics* **19**, 171-+ (2023). https://doi.org/10.1038/s41567-022-01849-9

22	Sie, E. J. *et al.* Valley-selective optical Stark effect in monolayer WS2. *Nature Materials* **14**, 290-294 (2015).

23	Ito, S. *et al.* Build-up and dephasing of Floquet-Bloch bands on subcycle timescales. *NATURE* **616**, 696-+ (2023). https://doi.org/10.1038/s41586-023-05850-x

24	Park, S. *et al.* Steady Floquet-Andreev states in graphene Josephson junctions. *Nature* **603**, 421-+ (2022). https://doi.org/10.1038/s41586-021-04364-8

25  Haxell, D. Z. et al. Microwave-induced conductance replicas in hybrid Josephson junctions without Floquet-Andreev states. *Nature communications* **14**, 6798 (2023). https://doi.org/10.1038/s41467-023-42357-5
26  KOMINAMI, M., POZAR, D. & SCHAUBERT, D. DIPOLE AND SLOT ELEMENTS AND ARRAYS ON SEMI-INFINITE SUBSTRATES. *IEEE TRANSACTIONS ON ANTENNAS AND PROPAGATION* **33**, 600-607 (1985).
27  Biagioni, P., Huang, J., Duò, L., Finazzi, M. & Hecht, B. Cross Resonant Optical Antenna. *PHYSICAL REVIEW LETTERS* **102** (2009). https://doi.org/10.1103/PhysRevLett.102.256801
28  Calvo, H. L., Pastawski, H. M., Roche, S. & Torres, L. Tuning laser-induced band gaps in graphene. *Applied Physics Letters* **98** (2011). https://doi.org/10.1063/1.3597412
29  Cong, X. et al. Probing the acoustic phonon dispersion and sound velocity of graphene by Raman spectroscopy. *CARBON* **149**, 19-24 (2019). https://doi.org/10.1016/j.carbon.2019.04.006
30  Nordhagen, E., Sveinsson, H. & Malthe-Sorenssen, A. Diffusion-Driven Frictional Aging in Silicon Carbide. *TRIBOLOGY LETTERS* **71** (2023). https://doi.org/10.1007/s11249-023-01762-z
31  Seetharam, K. I., Bardyn, C. E., Lindner, N. H., Rudner, M. S. & Refael, G. Controlled Population of Floquet-Bloch States via Coupling to Bose and Fermi Baths. *Physical Review X* **5** (2015). https://doi.org/10.1103/PhysRevX.5.041050
32  Seetharam, K., Bardyn, C., Lindner, N., Rudner, M. & Refael, G. Steady states of interacting Floquet insulators. *PHYSICAL REVIEW B* **99** (2019). https://doi.org/10.1103/PhysRevB.99.014307
33  Esin, I., Rudner, M., Refael, G. & Lindner, N. Quantized transport and steady states of Floquet topological insulators. *PHYSICAL REVIEW B* **97** (2018). https://doi.org/10.1103/PhysRevB.97.245401
34  Yang, C., Esin, I., Lewandowski, C. & Refael, G. Optical Control of Slow Topological Electrons in Moire Systems. *PHYSICAL REVIEW LETTERS* **131** (2023). https://doi.org/10.1103/PhysRevLett.131.026901
35  Li, Q., Hwang, E. & Das Sarma, S. Disorder-induced temperature-dependent transport in graphene: Puddles, impurities, activation, and diffusion. *PHYSICAL REVIEW B* **84** (2011). https://doi.org/10.1103/PhysRevB.84.115442
36  Esin, I., Gupta, G., Berg, E., Rudner, M. & Lindner, N. Electronic Floquet gyro-liquid crystal. *NATURE COMMUNICATIONS* **12** (2021). https://doi.org/10.1038/s41467-021-25511-9
37  Katz, O., Refael, G. & Lindner, N. Optically induced flat bands in twisted bilayer graphene. *PHYSICAL REVIEW B* **102** (2020). https://doi.org/10.1103/PhysRevB.102.155123
38  Li, Y., Fertig, H. & Seradjeh, B. Floquet-engineered topological flat bands in irradiated twisted bilayer graphene. *PHYSICAL REVIEW RESEARCH* **2** (2020). https://doi.org/10.1103/PhysRevResearch.2.043275
39  Kalugin, N. G. et al. Photoelectric polarization-sensitive broadband photoresponse from interface junction states in graphene. *2d Materials* **4** (2017). https://doi.org/10.1088/2053-1583/4/1/015002
40  Yang, Y. F. et al. Low Carrier Density Epitaxial Graphene Devices On SiC. *Small* **11**, 90-95 (2015). https://doi.org/10.1002/smll.201400989

41  El Fatimy, A. *et al.* Epitaxial graphene quantum dots for high-performance terahertz bolometers. *Nature Nanotechnology* **11**, 335-+ (2016). https://doi.org/10.1038/nnano.2015.303

**Extended data:**

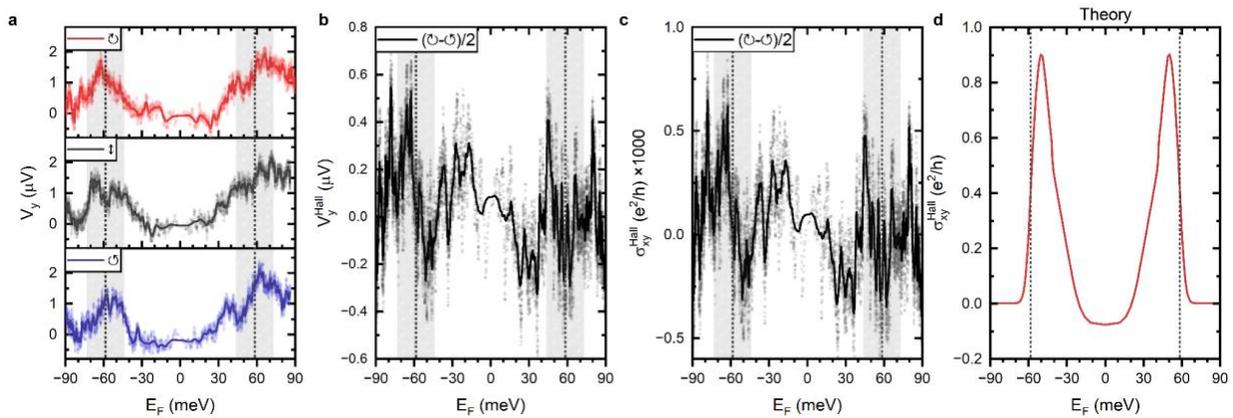

**Figure E1. Transverse voltage response under different conditions. a** Transverse voltage as a function of $E_F$ under light irradiation ($\hbar\Omega$ = 117 meV) with different polarizations of the laser. **b** helicity-dependent Hall voltage in Sample A. **c** Hall conductivity in Sample A. Here, the laser intensity is 1.4 mW/μm², bias source-drain voltage is 6 mV, and cryostat temperature is 3.4 K **d** Theoretically calculated Floquet Hall conductivity with nontrivial $E_F$-dependence arising from the tunable electronic steady state.

# Supplementary Information for: "Evidence of Floquet electronic steady states in graphene under continuous-wave mid-infrared irradiation"

Yijing Liu[1,#], Christopher Yang[2,#], Gabriel Gaertner[3], John Huckabee[3], Alexey V. Suslov[4], Gil Refael[2], Frederik Nathan[5], Cyprian Lewandowski[4,6], Luis E. F. Foa Torres[7], Iliya Esin[2,8,*], Paola Barbara[1,*], Nikolai G. Kalugin[3,*].

[1] Department of Physics, Georgetown University, Washington, DC 20057, USA.

[2] Department of Physics, IQIM, California Institute of Technology, Pasadena, California 91125, USA.

[3] Department of Materials and Metallurgical Engineering, New Mexico Tech., Socorro, NM 87801, USA.

[4] National High Magnetic Field Laboratory, Tallahassee, Florida 32310, USA

[5] Niels Bohr Institute, University of Copenhagen, Copenhagen 2100, Denmark

[6] Department of Physics, Florida State University, Tallahassee, Florida 32306, USA

[7] Department of Physics, Faculty of Physical and Mathematical Sciences, University of Chile, Santiago 837045, Chile

[8] Department of Physics, Bar-Ilan University, 52900, Ramat Gan, Israel

# These authors contributed equally

*e-mail: nikolai.kalugin@nmt.edu, paola.barbara@georgetown.edu, iesin@caltech.edu.

# Supplementary information

## A. Sample irradiation with continuous wave mid-infrared laser beam

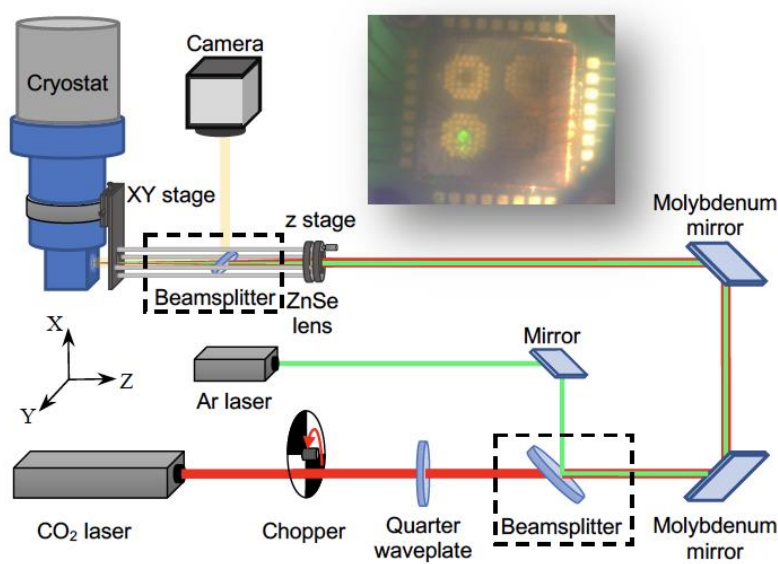

**Figure S1. Scheme of the sample illumination system.** Schematic illustration of the optical setup used for delivering and focusing the mid-infrared irradiation. The beamsplitters are used only during the preliminary alignment of optical elements. The inset image shows the visible laser spot on the sample used for the preliminary alignment of the molybdenum mirrors and the ZnSe focusing lens.

The optical setup, with the components described in the Materials and Methods section, is shown in Figure S1. A pair of molybdenum mid-infrared (Mid-IR) high-power mirrors were used for the delivery of the laser beam. For the pre-alignment of the orientation of the molybdenum mirrors and for the initial positioning of ZnSe lens closer to the position in focus, we used a visible laser ($\lambda = 520$ nm) collimated with the Mid-IR irradiation, with a camera to monitor the laser spot on the sample (see the inset in Figure S1). After this initial alignment, the visible laser and the beamsplitters were removed, and the fine adjustment of the beam positioning with the Mid-IR laser was performed by using the photo-generated current in our devices as a reference. The illumination caused a change in the source-drain current due to the photo-generated hot electrons, as described in the main text. An example of current change under 10.6-μm-wavelength laser illumination is Fig. S2 below. We adjust the beam position by maximizing the photocurrent response.

The devices were irradiated with a 10.6-μm-wavelength CW $CO_2$ laser, providing up to 25 W of linearly polarized light. The laser beam polarization was controlled with a λ/4 plate. Experimental samples were designed to maximize efficiency of the beam-sample coupling for the mid-IR irradiation[1]. At the same time, the half-wavelength size of the graphene devices and the attachment of the metallic electrodes to its edges result in predominantly linear polarization of the incident electromagnetic field near the edges of the graphene sample, regardless of the initial polarization of the laser source.[2] The presence of linearly polarized fields in some parts of the sample is expected to influence helicity-dependent effects.

**B. Measurement sequence**

The laser beam was modulated by a chopper (the modulation frequencies were in the range of 10-50 Hz) with a duty cycle of ~10 %. During the on period, the system forms a non-equilibrium electronic state characterized by a photocurrent $I_{dr}$ that relaxes during the off period to an equilibrium state with temperature $T_{eq}$ and corresponding photocurrent $I_{eq}(T_{eq})$. The relaxation occurs over the phonon scattering time, typically on the picosecond scale. The measured photocurrent is, therefore, given by $\Delta I = |I_{dr} - I_{eq}(T_{eq})|$, which is detected directly using a lock-in amplifier, with the reference signal to the lock-in provided by a chopper. Figure S2 shows the theoretically predicted photocurrent as a function of time, switching between $I_{dr}$ and $I_{eq}(T_{eq})$, within the duty cycle.

Due to the illumination, the system temperature $T_{eq}$ is elevated above the background temperature $T_{bg}$. To test the temperature change and the corresponding bolometric effect, we blocked the beam in the optical path between the chopper and the sample for extended times of a few seconds. During a timescale denoted $\tau_r$, the system's temperature relaxed back to $T_{bg}$, corresponding a reduction of the current to the value $I_{eq}(T_{bg})$. We note that the background itself is slightly heated by the laser, leading to gradual changes in $T_{bg}$. The bottom panel of Figure S2 shows the longitudinal (source-drain) current measured as a function of time and at constant source-drain voltage while the laser beam was blocked or unblocked. The time dependence of the current can be fit to the function $I_{eq}(T(t))$, where $T(t) = T_{bg} + (T_{eq} - T_{bg})e^{-t/\tau_r}$ within one unblocking-blocking interval.

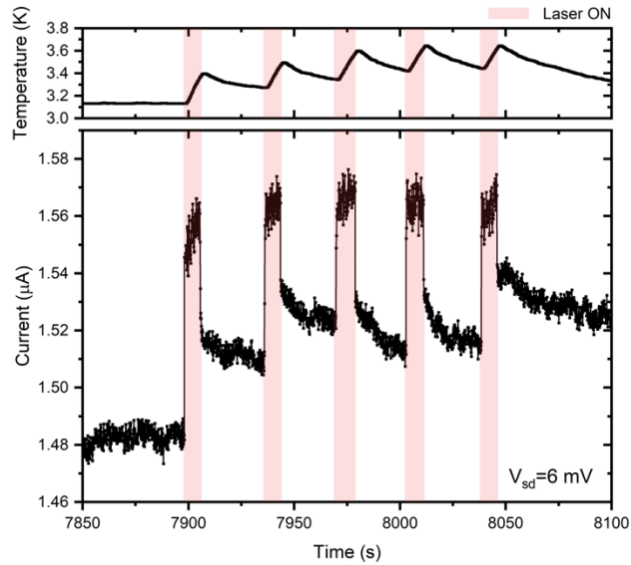

**Figure S2. Measurement sequence in time.** Experimentally measured cryostat temperature (top panel) and longitudinal current (bottom panel) with laser blocked and unblocked intervals. Note that the actual graphene lattice temperature, $T(t)$, is expected to be further elevated relative to the cryostat temperature. When the laser is blocked, the sample cools down to $T_{bg}$, giving rise to a strong bolometric effect in the longitudinal current. The laser illumination is blocked/unblocked in the optical path between the chopper and the sample (grey stripes indicate when laser is unblocked), the average laser power density is 0.125 mW/µm² and the peak power density is 1.25 mW/µm², $\hbar\Omega$ = 117 meV (10.6 µm wavelength).

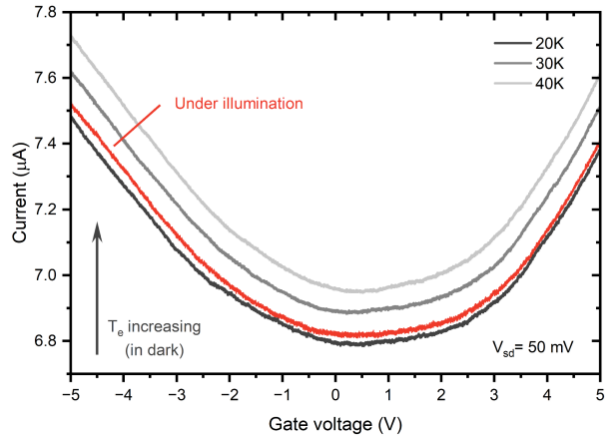

**Figure S3.** Laser intensity and temperature dependence of the longitudinal conductance. Gate voltage dependence of the source-drain current at different temperatures with laser blocked (gray) and under irradiation at the base temperature (red), with $P = 0.6$ mW/μm$^2$ and with source-drain voltage fixed at 50 mV.

The relation of the photoresponse $\Delta I$ to effects of electron heating may be illustrated further in the following supplementary experiment. Figure S3 shows the current vs. gate voltage curves of a graphene device at three different temperatures and at a fixed source-drain bias voltage without any laser irradiation (gray curves). The temperature dependence of the electrical resistance in graphene is determined by different intrinsic and extrinsic sources of scattering, including phonons, defects in the crystal lattice or deformation in the graphene sheet (wrinkles or steps), as well as impurities in the substrate or on the graphene surface.[3-7] The presence of electron-hole puddles and quantum corrections also complicate the dependence of the electrical resistance on carrier density and temperature, especially at low temperature.[8-11] For our graphene samples, in the whole range of gate voltage that we measured around the Dirac point, the current increases when the temperature increases.

**C. Transmission property of ITO top gate for Mid-IR irradiation**

The Mid-IR radiation actually delivered to the graphene is attenuated by the optical elements. Our ZnSe focusing lens has an anti-reflective (AR) coating, while the cryostat window has no AR coating and transmits 70% of the 10-μm wavelength laser beam.

The other important source of attenuation is the Indium-Tin Oxide (ITO) material of the top gate in the experimental samples. Depending on growth conditions, post-treatment, and many other factors, the optical transmission of this material may vary significantly. Measured or theoretically expected values of Mid-IR transmission of ITO range from less than 10% to more than 50% for material thicknesses around 100 nm, depending on the electron concentrations.[12,13] For an accurate estimation of attenuation in our case, we fabricated and tested ITO samples using the exact same growth conditions and thickness of the material as those we used in our experimental graphene devices. The infrared transmission spectra were measured using a Nicolet Nexus 8700 Fourier-Transform spectrometer equipped with a KBr beamsplitter and a DTGS photodetector. A typical transmission spectrum of a 110-nm-thick ITO layer is shown in Figure S2. Transmission measurements were performed at multiple locations on two different samples, yielding an average transmission of 45%.

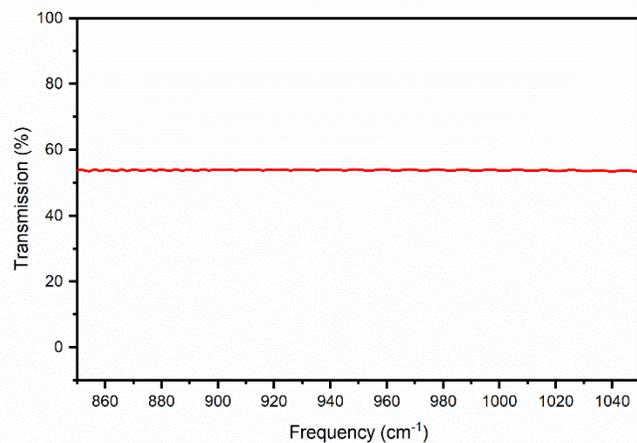

**Figure S4. Mid-Infrared transmission spectrum of ITO.** Example of spectrum measured from our typical 110-nm thick ITO layer in the range around 10.6 μm wavelength (corresponding to the wavenumber 943 cm$^{-1}$).

## D. Cyclotron resonance of the top-gated epitaxial graphene samples

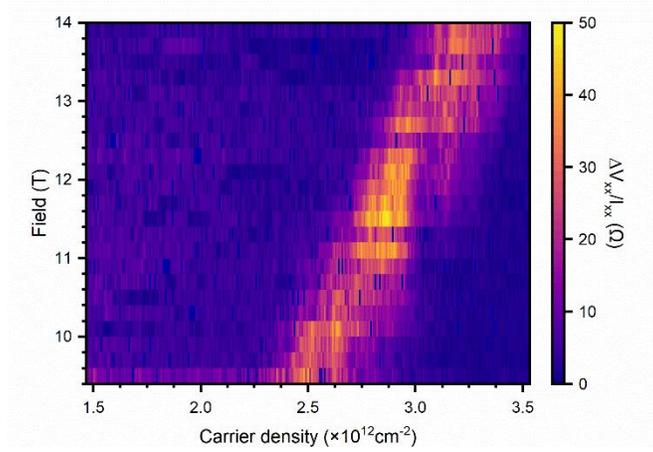

**Figure S5. Longitudinal photoresponse vs. carrier density at different B fields.** Longitudinal voltage photoresponse as a function of magnetic field and carrier density, under chopper-modulated irradiation with circular polarization, at a laser intensity of 20 µW/µm² and photon energy $\hbar\Omega = 117$ meV. The sample is biased at a fixed current $I_x$ of 1 µA, and the photoinduced change in source-drain voltage $\Delta V_x$ is measured with a lock-in amplifier.

We performed cyclotron resonance measurements on an epitaxial graphene sample on SiC at 1.6 K using an 18/20 T general purpose superconducting magnet at NHMFL in Tallahassee, FL. We used a 1 MOhm resistor connected in series with the sample to keep the source-drain current at a constant value of about 1 µA throughout the measurements. The sample was illuminated by a CW $CO_2$ laser and a quarter waveplate, as described above. Unlike the zero-field measurement, the laser was delivered through a Mid-IR fiber and focused onto the sample with a ZnSe lens. As shown in Figure S5, the magnetoresistance reveals a resonance magnetic field value of $11.5 \pm 0.2$ T that corresponds to a transition between the 0th and the 1st Landau levels, or

$$\Delta E = v_F\sqrt{2\hbar eB/c} = \hbar\Omega,$$

where $\hbar\Omega = 117$ meV and $B = 11.5 \pm 0.2$ T. We can, therefore, extract the Fermi velocity in our epitaxial graphene as $v_F = (0.960 \pm 0.008) \times 10^6$ m/s. The cyclotron resonance peak is determined to have an FWHM of ~8 meV, which is about two times broader than that reported in exfoliated graphene samples.[14,15] Such broadening can be attributed to the charge puddles and traps in graphene introduced during the fabrication process.

We can also extract the magnetic length at the resonance field $l_B$ ~7.6 nm, which is smaller than our estimated mean free path ($l_{MFP}$) for carrier density $n_c$ ~2.9×10¹² cm⁻², about $l_{MFP}$~30 nm.

### E. Gate efficiency calibration

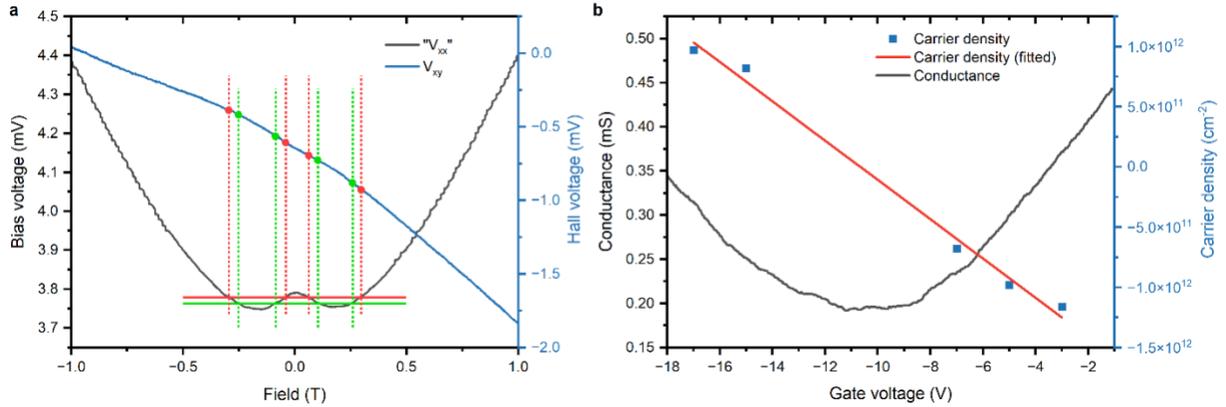

**Figure S6. Gate efficiency calibration. a** (Black) Source-drain voltage and (blue) transverse voltage from the Hall effect measurements for $V_g = 7$ V and a constant bias (source-drain) current $I_{sd} = 1$ µA. The green and red lines are examples of the set of "4 points" used for extracting the actual transverse voltage. **b** (Black) Conductance at different gate voltages. (Blue) Carrier density at different $V_g$ from Hall measurements. (Red) Linear fitting of carrier density at different $V_g$.

The Hall effect measurements were performed on sample B at 4 K to determine the carrier density at different gate voltages using a Quantum Design PPMS® system. A 1 MOhm resistor was connected in series with the device to keep the current around 1 µA throughout the measurements. The typical Hall voltage and source-drain voltage curves as a function of the magnetic field at a fixed gate voltage are plotted in Figure S6 **a**, where the Hall voltage revealed a non-linear behavior as well as an offset. We note that these source-drain voltages are two-terminal measurements. Unlike four-terminal measurements, two-terminal measurements yield curves that depend on the geometry of the samples.[16] In addition, a slight misalignment of the Hall terminals may lead to a purely geometric (not Hall-effect-related) contribution to the Hall voltage from the bias current that is proportional to the source-drain voltage. We estimated this misalignment as ~300 nm along the source-drain current direction in the presented sample. To exclude the contribution from such misalignment, we used the "four-point" approach described below to extract the linear relationship between the Hall voltage and the applied magnetic field: we randomly picked four data points around the bottom of the source-drain voltage curve at different fields, but with the same source-drain voltage value, so that the misalignment would contribute equally to the Hall voltages among these four points, and we extracted the carrier density from a linear fitting of those four points. For each gate voltage, three sets of the "four points" were randomly picked. Figure S6 **a** shows two sets of four points, in red and green, respectively, for measurements at a fixed gate voltage $V_g = -7$ V. For each set of 4 points the following linear fitting was performed:

$$V_{\text{Hall}}[\text{V}] = k[\text{VT}^{-1}]B[\text{T}] + V_{\text{Hall offset}}[\text{V}],$$

where the slope $k$ links to the charge carrier density $n_c$ with the following relationship:

$$n_c[\text{cm}^{-2}] = 10^{-4} \times \frac{I[\text{A}]}{ek[\text{VT}^{-1}]},$$

where $I$ was kept around 1 μA as mentioned above and $e = 1.60 \times 10^{-19}$ C. Table **S1** summarizes the data obtained from the fittings and the carrier concentrations obtained at different gate voltages. The extracted values of carrier concentrations at different gate voltages are plotted in Figure S6 **b**. From a linear fit of those points, we obtained the following "calibration curve" to relate the carrier density to the gate voltage:

$$n_c[\text{cm}^{-2}] = -k(V_g - V_d)[\text{V}]$$

where $k = 1.63 \times 10^{11} \pm \delta k$, $\delta k = 9 \times 10^9$, $V_d = kn_0 = 10.7$ V $\pm \delta V_d$, and $\delta V_d \approx V_d\sqrt{(\delta n_0/n_0)^2 + (\delta k/k)^2}$. Here, $n_0 = 1.74 \times 10^{12}$, and $\delta n_0 = 9.37 \times 10^{10}$, with the uncertainties estimated from the 68% confidence interval of the linear fit. The following relationship between the carrier density $n_c$ and Fermi energy $E_F$ in graphene was then applied to determine the corresponding $V_g$ for $E_F = \pm 1/2\,\hbar\Omega$:

$$E_F = \hbar v_F \sqrt{\pi n_c},$$

where $\hbar$ is the reduced Planck constant, $v_F$ is the Fermi velocity extracted from the cyclotron resonance measurements as $0.96 \times 10^6$ m/s in our top-gated epitaxial graphene samples. For the first crossings, the Fermi energy is half the photon energy, i.e., $E_F = \pm 58.5$ meV, and the corresponding gate voltage was calculated to be about 1.6 V from the charge neutrality point, as summarized in Table **S2**.

**Table S1.** Fitting results of Hall measurements at different gate voltages

|  |  | Gate voltage (V) | | | | |
|---|---|---|---|---|---|---|
|  |  | -3V | -5V | -7V | -15V | -17V |
| Slope (×10⁻⁴) /VT⁻¹ | Set 1 | -5.290 | -6.244 | -9.074 | 7.458 | 6.294 |
|  | Set 2 | -5.250 | -6.247 | -9.024 | 7.521 | 6.369 |
|  | Set 3 | -5.283 | -6.267 | -9.029 | 7.412 | 6.281 |
| Ave. slope (×10⁻⁴)/ VT⁻¹ | | -5.274 | -6.252 | -9.042 | 7.464 | 6.315 |

**Table S2.** $V_g$ corresponds to $E_F = \pm 1/2\ \hbar\Omega$

|  | $-1/2\hbar\Omega$ | CNP | $+1/2\hbar\Omega$ |
|---|---|---|---|
| $E_F$/meV | −58.5 | 0 | +58.5 |
| $n_c$/cm² | −2.51 × 10¹¹ | 0 | +2.51 × 10¹¹ |
| $V_g$/V | −12.2 | −10.7 | −9.1 |

## F. Device Fabrication

The top-gated graphene FETs were fabricated using epitaxial graphene on SiC purchased from Graphene Waves. To prevent sample contamination from photoresists, a thin Pd/Au layer (5 nm Pd+15 nm Au) was deposited on the graphene by electron-beam evaporation before further processing.[17,18] The contacts were fabricated first to prevent the charging effect during the electron beam lithography (EBL) process. Contact patterns were written following a standard EBL process using a Zeiss SUPRA55-VP system on a methyl methacrylate/polymethyl methacrylate (MMA/PMMA) bilayer e-beam resist, followed by magnetron sputtering deposition of a total of 1.5 nm Ti and 400 nm Au. The graphene was then patterned into Hall bars following a process described in Figure **S7 a**, where a layer of PMMA defined by EBL was used as the etch mask during dry etching (Ar plasma, 50 s.c.c.m., 150 W, Oxford Plasmalab 80), and the EBL pattern was designed in such a way that only the channel area shown in Figure **S7 b** remained unexposed. Prior to the top gate fabrication, diluted aqua regia (DAR, $HNO_3:HCl:H_2O$ = 1:3:4) was used to remove the Pd/Au protection layer. The samples were annealed in vacuum afterwards to remove the residues from the DAR treatment as well as adsorbates from ambient exposure ($H_2O$, $O_2$, etc.).

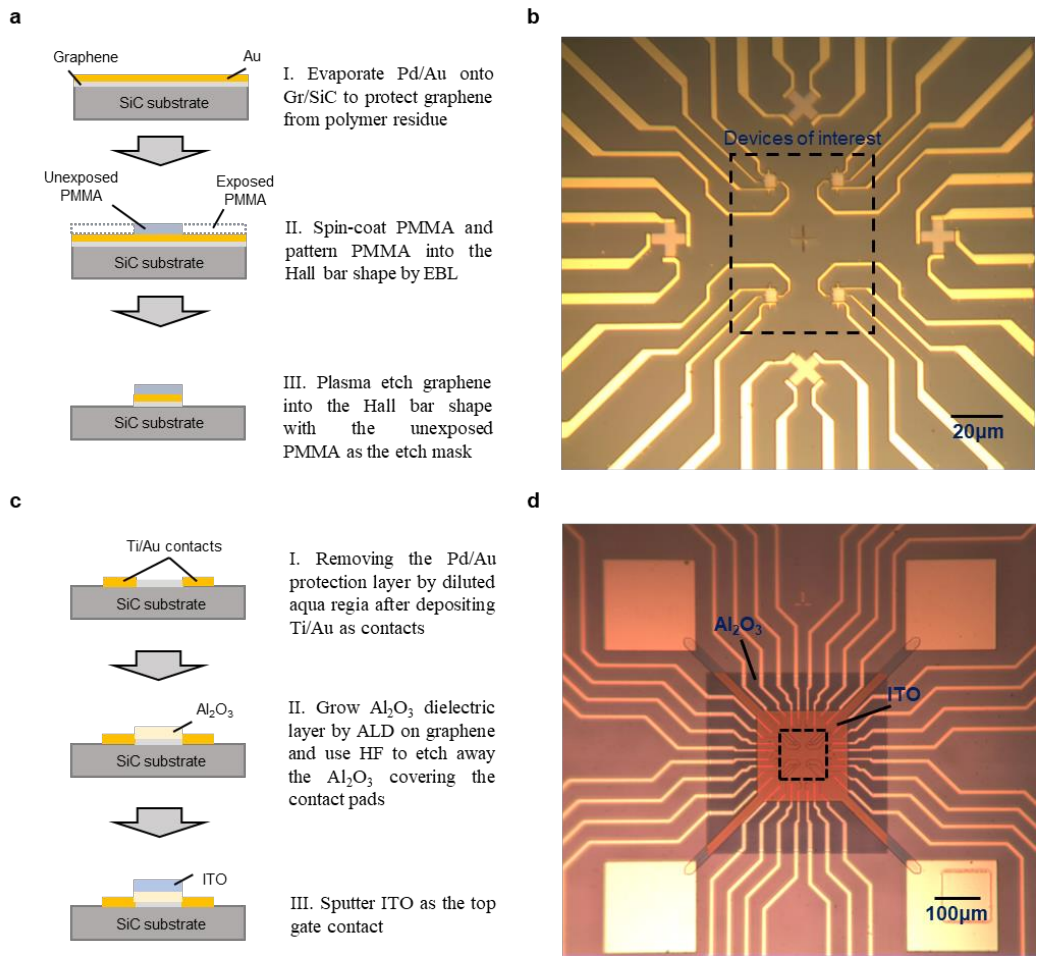

**Figure S7. Scheme of key steps in top-gated graphene device fabrication. a** Patterning of graphene Hall bars with a thin Pd/Au protection layer. **b** Optical image of the graphene devices prior to removing the metal protection layer and the top gate fabrication. **c** Illustration of top gate fabrication using $Al_2O_3$ as the dielectric layer and ITO as the gate contact. **d** Optical image of the top-gated devices. The dashed line marks the area of devices shown in **b**.

Since the graphene grown on SiC is intrinsically n-doped due to the formation of the buffer layer during the graphene growth, before the atomic layer deposition (ALD) growth of an $Al_2O_3$ layer, we doped the samples with nitric acid vapors in a standard fume hood, following a procedure developed by *Mhatre et al.*[19] to bring the Fermi level closer to the charge neutrality point.

As shown in Figure **S7 c**, a 90-nm $Al_2O_3$ dielectric layer was grown on top of graphene by atomic layer deposition (Beneq TFS 200). The $Al_2O_3$ masking of the contact pads was removed by buffered hydrofluoric acid using a layer of Shipley 1813 patterned by photolithography as the etch mask. The top gate contact was patterned by photolithography, followed by sputtering of 110-nm indium tin oxide (ITO). Figure **S7 b** shows a cluster of four top-gated devices.

## G. Characterization of single-particle dynamics via Floquet states

We first derive the driven graphene Hamiltonian near the valleys $\xi = +1, -1$ corresponding to the $K$ and $K'$ valleys of graphene. The Dirac Hamiltonian in valley $\xi$ is given by $H^\xi(k) = \hbar v_F \vec{k} \cdot (\sigma_x, \xi \sigma_y)$, where $\vec{k} = (k_x, k_y)$ is the electronic momentum and $\sigma_x$ and $\sigma_y$ are the Pauli matrices. We include the irradiation by the circularly polarized laser of vector potential $\vec{A}(t) = A(\cos \Omega t, \sin \Omega t)$ via minimal coupling: $H^\xi(\vec{k}, t) = H^\xi(\vec{k} + e\vec{A}(t)/\hbar)$. The time evolution generated by $H^\xi(\vec{k}, t)$ has a complete set of solutions of the form $|\psi_{\vec{k}\alpha}^\xi(t)\rangle = e^{-i\varepsilon_{\vec{k}\alpha} t/\hbar} |\Phi_{\vec{k}\alpha}^\xi(t)\rangle$, where $|\Phi_{\vec{k}\alpha}^\xi(t)\rangle = |\Phi_{\vec{k}\alpha}^\xi(t + T)\rangle$ is the Floquet-Bloch state, $\varepsilon_{k\alpha}$ is the quasienergy, and $T = 2\pi/\Omega$ denotes the laser driving period. (Here, we drop the $\xi$ index in $\varepsilon_{k\alpha}$, because the quasienergy spectrum is identical in both valleys.) As a result of their time periodicity, the Floquet-Bloch states can be decomposed as $|\Phi_{\vec{k}\alpha}^\xi(t)\rangle = \sum_m e^{-im\Omega t} |\phi^{n\xi}_{\vec{k}\alpha}\rangle$, with $|\phi^{n\xi}_{\vec{k}\alpha}\rangle$ denoting the Floquet harmonics that can be found by solving the Floquet Schrodinger equation (see Ref. [20]). The Floquet-Bloch states form a complete basis of stationary solutions to the time-evolution generated by $H^\xi(\vec{k}, t)$, which are taken onto themselves after each driving period, up to a phase controlled by the quasienergy $\varepsilon_{k\alpha}$. Note that $\varepsilon_{k\alpha}$ is only defined up to an integer multiple of the photon energy $\hbar\Omega$, compensated by a redefinition of $|\Phi_{\vec{k}\alpha}^\xi(t)\rangle$ via multiplication by factor(s) of $e^{i\Omega t}$. We use the convention where $\varepsilon_{k\alpha}$ is chosen such that $\max_n |\langle \phi^{n\xi}_{\vec{k}\alpha} | \phi^{n\xi}_{\vec{k}\alpha} \rangle| = |\langle \phi^{0\xi}_{\vec{k}\alpha} | \phi^{0\xi}_{\vec{k}\alpha} \rangle|$.

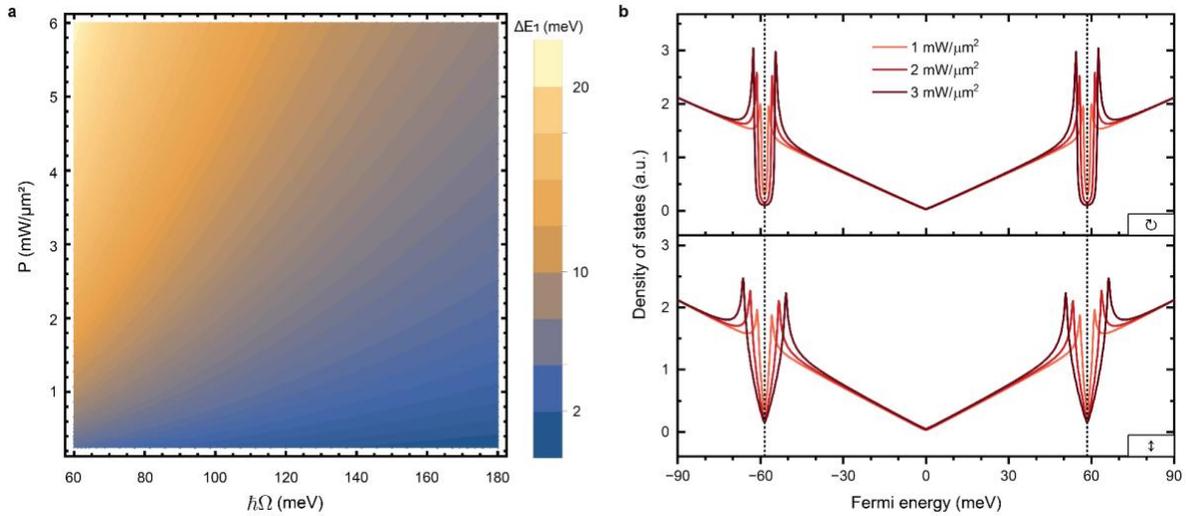

**Figure S8. Characteristics of the single-particle quasienergy spectrum. a Single-particle quasienergy** bandgap $\Delta E_1$ opening at $\varepsilon_{k\alpha} = \pm 1/2\ \hbar\Omega$, with $\varepsilon_{k\alpha} = 0$ corresponding to the Dirac point. The axes indicate the laser power density and photon energy. **b** Predicted light-induced modifications of the time-averaged density of states of graphene (see section J for definition) as a function of Fermi energy shift from the Dirac point for different power densities, with circular polarization (top) and linear polarization (bottom). Notice that the laser-induced gap at the Dirac point, which is a second-order process, is not resolved in the experiment for the chosen laser intensities. (See the main text for details.)

Laser illumination with a circularly polarized laser opens dynamic gaps in the Floquet quasienergy spectrum. The gap opening at the resonance energies around $\varepsilon_{k\alpha} = \pm 1/2\ \hbar\Omega$, with a size $\Delta E_1$ linear in the laser field amplitude, controls the Rabi frequency of the resonant inter-band transitions induced by the laser drive. At the Dirac point ($\varepsilon_{k\alpha} = 0$), the band degeneracy is lifted by a second-order resonance coupling between the conduction and valence bands, with a gap opening quadratic in the laser field amplitude.[21-23] In the low power regime, $\Delta E_1$ is, therefore, larger than the bandgap at the Dirac point. Illustrating this, Figure S8 **a** shows the theoretical prediction for the quasienergy bandgap $\Delta E_1$ as a function of laser power density and photon energy. For the photon energy and power density used in our experiment, $\Delta E_1$ is predicted to be in the range of ~ 1-10 meV, while the gap around the Dirac point, too small to be resolved in our measurements, cannot be clearly discerned in the numeric calculations.

## H. Steady state simulations of laser-illuminated graphene

To determine the non-equilibrium steady state occupation $F_{\vec{k}\alpha}^{\xi}$ of the Floquet bands,[24-27] we consider the Floquet-Boltzmann equation (FBE) for electron-phonon collisions, which describes the flow of electronic occupations into Floquet states. The FBE is given by $\partial_t F_{\vec{k}\alpha}^{\xi} = I^{\text{el-ph}}{}_{\vec{k}\alpha}^{\xi}[\{F_{\vec{k}\alpha}^{\xi}\}]$, where $I^{\text{el-ph}}{}_{\vec{k}\alpha}^{\xi}[\{F_{\vec{k}\alpha}^{\xi}\}]$ is the electron-phonon collision integral. To estimate the collision integral, we consider electronic coupling to graphene-SiC surface acoustic phonons and graphene longitudinal acoustic phonons, indexed by $j = 0$ and $1$, respectively. Assuming weak scattering rates relative to the Floquet gap $\Delta E_1$, we use Fermi's golden rule to calculate the electron-phonon collision integral

$$I^{\text{el-ph}}{}_{\vec{k}\alpha}^{\xi}[\{F_{\vec{k}\alpha}^{\xi}\}] \approx \frac{1}{N}\frac{2\pi}{\hbar}\sum_{\xi'=\pm 1}\sum_{\vec{k}'\in\text{BZ}}\sum_{\alpha'}\sum_j\sum_n |G^{\vec{k}'\alpha'\xi'}{}_{\vec{k}\alpha\xi}(n,j)|^2 \times$$

$$\times [\{F_{\vec{k}'\alpha'}^{\xi'}(1 - F_{\vec{k}\alpha}^{\xi})\mathcal{N}(\hbar\omega_j(\vec{k}'-\vec{k})) - F_{\vec{k}\alpha}^{\xi}(1 - F_{\vec{k}'\alpha'}^{\xi'})[1 + \mathcal{N}(\hbar\omega_j(\vec{k}'-\vec{k}))]\}$$

$$\times p(\varepsilon_{\vec{k}'\alpha'} - \varepsilon_{\vec{k}\alpha} + \hbar\omega_j(\vec{k}'-\vec{k}) + n\hbar\Omega)$$

$$+\{F_{\vec{k}'\alpha'}^{\xi'}(1 - F_{\vec{k}\alpha})[1 + \mathcal{N}(\hbar\omega_j(\vec{k}'-\vec{k}))] - F_{\vec{k}\alpha}(1 - F_{\vec{k}'\alpha'}^{\xi'})\mathcal{N}(\hbar\omega_j(\vec{k}'-\vec{k}))\}$$

$$\times p(\varepsilon_{\vec{k}'\alpha'} - \varepsilon_{\vec{k}\alpha} - \hbar\omega_j(\vec{k}'-\vec{k}) + n\hbar\Omega)],$$

where $\omega_j(\vec{k}'-\vec{k}) = c_{\text{ph}}^j|\vec{k}'-\vec{k}|$ is frequency dispersion of the phonon branch $j$, $c_{\text{ph}}^j$ is the speed of sound, $N$ is the number of discretized $\vec{k}$-points on the 2D grid, and $\mathcal{N}(\varepsilon) = (e^{-\varepsilon/k_B T} - 1)^{-1}$ is the occupation of a phonon mode with energy $\varepsilon$, evaluated at temperature $T$. The matrix element

$$G^{\vec{k}'\alpha'\xi'}{}_{\vec{k}\alpha\xi}(n,j) = M_{j,\vec{k},\vec{k}'}\sum_m\sum_{\nu\nu'}\langle\phi^{n+m,\xi}{}_{\vec{k}\alpha}|\phi^{m\xi}{}_{\vec{k}\alpha}\rangle,$$

describes the electronic coupling to the $j$-th phonon branch, where

$$M_{j,\vec{k},\vec{k}'} = \frac{1}{\sqrt{A}}\frac{D^j}{\sqrt{2\rho c_{\text{ph}}^j}}\sqrt{\hbar\omega_j(\vec{k}'-\vec{k})},$$

$A$ is the unit cell size of graphene, $\rho$ is the density of graphene, and $D^j$ is the deformation potential corresponding to the phonon branch $j$. To impose energy conservation on electron-phonon scattering processes, we use the smeared Dirac-Delta function

$$p(\varepsilon) = 1.05 e^{-\varepsilon^2/2\sigma^2}/(2.51\sigma)\theta(2\sigma - |\varepsilon|),$$

where $\theta(x)$ is the Heaviside step function and the numerical prefactors ensure normalization ($\int_{-\infty}^{\infty} p(\varepsilon)d\varepsilon = 1$). Here, the phenomenological level broadening of $\sigma = 3$ meV describes the phonon and electronic spectral broadening due to disorder in the system. Our calculations assume that the speeds of the phonon branches are given by $c_{ph}^0 = 1.3$ km/s and $c_{ph}^1 = 11$ km/s and their deformation potentials satisfy $D^0/D^1 = 0.18$.

To obtain the steady state occupations, we solve for $I^{el-ph}_{\vec{k}\alpha}{}^{\xi}[\{F_{\vec{k}\alpha}{}^{\xi}\}] = 0$ using the Newton-Raphson method. Our calculations are performed on a discretized square momentum grid covering the region $k_x, k_y \in [-0.17 \text{ nm}^{-1}, 0.17 \text{ nm}^{-1}]$ with 105 equally spaced grid points in the $k_x$ and $k_y$ directions. We focus only on low-energy electronic states, where the Floquet occupations can be approximated as rotationally symmetric about the Dirac node, i.e., $F_{\vec{k}\alpha}{}^{\xi} \approx F_{k\alpha}$, with $F_{k\alpha}$ depending only on the momentum magnitude $k = |\vec{k}|$ and band index $\alpha$. To impose the electronic particle number, we add a Lagrange multiplier term $\lambda(\sum_{\vec{k}\alpha} F_{\vec{k}\alpha} - gN)/N$ to the FBE with a large constant $\lambda$, and we vary the doping by changing $g$. The corresponding Fermi energies $E_F = \hbar v_F\sqrt{n/\pi}$ in equilibrium can be calculated from the electron density $n = \int d^2k/(2\pi)^2 F_{\vec{k}+}^{+1}$ as determined using the numerically obtained steady state.

## I. Transport in the Floquet Steady State

To calculate the longitudinal photocurrent $\Delta I$ from the steady state, we use linear response theory to estimate the longitudinal conductivity in the driven system:

$$\sigma_{xx} \approx e^2 \sum_\alpha \int \frac{dk}{2\pi} \int d\theta \, k\tau(k) v_x^2 \frac{dF_{k\alpha}}{d\varepsilon_{k\alpha}}$$

where $v_x = |\nabla_{\vec{k}} \varepsilon_{\vec{k}\alpha}| \cos\theta$ and $\tau(k) \propto k$ for relaxation processes dominated by scattering from charge impurities.[7] The longitudinal photocurrent is given by $\Delta I = (\sigma_{xx} - \sigma_{xx}{}^0)V_{sd}$, where $V_{sd}$ is the source-drain voltage and $\sigma_{xx}{}^0$ is the longitudinal conductivity in the equilibrium system, obtained by replacing $v_x = v_F \cos\theta$ and $F_{k\alpha}$ with the Fermi-Dirac distribution.

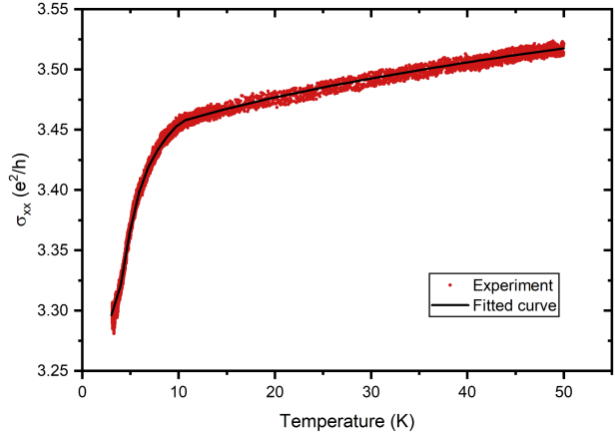

**Figure S9.** Temperature dependence of the longitudinal conductivity at $E_F = 0$ without irradiation, which indicates charge puddle effects.

In addition to the above-mentioned effects, charge puddles also modify the longitudinal conductivity. In particular, for doping near charge neutrality, where a clean graphene sheet normally exhibits low conductivity, the charge puddles introduce regions with larger chemical potential and greater mobility, thereby enhancing the conductivity.[28] To roughly capture this behavior, we employ a simple model to relate the conductivity $\sigma$ in clean graphene to that with charge puddles, $\sigma_p = \sigma + c(T)$, similar to that used in Ref.[6] Here, $c(T)$ is a function which captures the temperature-dependence of the experimentally-measured conductivity at charge neutrality, fit using $c(T) = \sum_{i=0}^{14} c_i T^i$ to the experimental data for $T < 10\ K$, and fit using the function $c(T) = d_1 + d_2 T^{1/2}$ for $T > 10\ K$, where $c_i$, $d_1$ and $d_2$ are fitting parameters (see Figure S9). Due to heating effects from the drive, the temperature during the duty cycle of the chopper differs when the laser is blocked by the chopper $T_{\text{eq}}$ and when the laser is unblocked $T_{\text{dr}}$. We account for the temperature mismatch and charge puddle effects by shifting the photocurrent by a magnitude $[c(T_{\text{dr}}) - c(T_{\text{eq}})]V_{\text{sd}}$ in our simulations. We find that this temperature mismatch accounts for the highly temperature and amplitude dependent photocurrent $\Delta I$ magnitude at charge neutrality $E_F = 0$ observed in the experiment. In particular, in our theory, we assume both $T_{\text{dr}} -$

$T_{eq} \approx 7.3 \text{ K/(mW/µm}^2)P$ and $T_{eq} = 3.5 \text{ K} + 1.83 \text{ K/(mW/µm}^2)P$ increase linearly as a function of the laser power density $P$. The coefficients chosen allow our theory to roughly capture the dependence of $\Delta I$ at charge neutrality as $T_{eq}$ and $P$ are varied in the experiment.

We emphasize that the fitting procedures for the variables $c(T)$, $T_{dr} - T_{eq}$, and $T_{eq}$ do not affect the width, depth, or shape of the photocurrent dip predicted in our theory; rather, it only shifts the $\Delta I$ by a constant independent of electronic doping.

### J. Density of States and Berry Curvature Calculation

The time-averaged density of states (defined as the time-dependent spectral function averaged over one period) was calculated numerically through the formula

$$\text{DOS}(\varepsilon) = \sum_{k\alpha} \delta(\varepsilon_{k\alpha} - \varepsilon)$$

where the broadening parameter $\sigma$ of the Gaussian function $\delta(\varepsilon)$ is chosen to be much smaller than the resonant Floquet gap.

The transverse conductivity was calculated from the anomalous conductivity in the steady state $\sigma_{xy} = \frac{2e^2}{\hbar} \sum_\alpha \int d^2\vec{k}\, B_{\vec{k}\alpha}{}^\xi F_{\vec{k}\alpha}{}^\xi$, where $B_{\vec{k}\alpha}{}^\xi$ is the time-averaged Berry curvature, given by[20]

$$B_{\vec{k}\alpha}{}^\xi = \frac{1}{2\pi/\Omega} \int_0^{2\pi/\Omega} \frac{(2\pi)^2}{N_b{}^2 A} \arg\left[\frac{U_x(\vec{k},t)U_y(\vec{k}+\hat{e}_x,t)}{U_x(\vec{k}+\hat{e}_y,t)U_y(\vec{k},t)}\right]$$

where

$$U_\mu(\vec{k},t) = \frac{\langle u_{\vec{k}\alpha}(t)|u_{\vec{k}+\hat{e}_\mu,\alpha}(t)\rangle}{\left|\langle u_{\vec{k}\alpha}(t)|u_{\vec{k}+\hat{e}_\mu,\alpha}(t)\rangle\right|^2},$$

$\hat{e}_\mu = 1/N_\mu \hat{u}$, $\mu = x, y$ and $|u_{\vec{k}\alpha}(t)\rangle$ are the Bloch vectors, defined by $|\psi_{\vec{k}\alpha}{}^\xi(t)\rangle = e^{i\vec{k}\cdot\vec{r}}|u_{\vec{k}\alpha}(t)\rangle$.

## K. Transverse Hall Voltage

An expected consequence of the Floquet states described above is the appearance of a Hall signal.[21,26,27] We measure the transverse voltage in samples under irradiation with both circular and linear polarization, as shown in extended Figure E1 **a** and in Figure S10. Floquet theory predicts that a nonzero transverse voltage originates only from nontrivial band topology, inconsistent with the observation of a transverse signal for linear polarization. However, the extraneous, helicity-independent transverse signal has been observed before under linearly-polarized Mid-IR irradiation at zero source-drain bias in McIver et al.[29] and attributed to an effective extra bias voltage caused by band bending fields at the contacts, a possible result of asymmetries introduced by charge puddles. The helicity-independent component may also be a product of photovoltaic effects generated by optical transitions in highly irradiated graphene [30,31] or anisotropies in the Floquet band spectrum near the resonant Floquet gap,[23] which may appear due to the distorted circular polarization near the metallic contacts (see Section A).

The weak polarization-dependent Hall voltage exhibits signs of a nontrivial $E_F$-dependence, shown in Figure E1 **b** for graphene sample A, displaying peaks starting to emerge around doping $E_F = \pm\hbar\Omega/2$. In Figure E1 **c**, we show the corresponding Hall conductivity. Upon inverting the sign of the source-drain bias $V_{sd}$, the Hall voltage peaks reverse sign. Such a helicity-dependent Hall voltage contribution may be an indicator of the emergence of Floquet chiral edge states in our samples. The theoretically calculated Floquet Hall conductivity, see Figure E1 **d**, also exhibits peaks near doping $E_F = \pm\hbar\Omega/2$. However, the magnitude of the Hall conductivity observed in the experiment is much weaker than the theory-predicted levels. There are several possible reasons for this discrepancy. One possible contribution to the suppressed Hall voltage is the distorted polarization of the laser beam near the metallic contacts (see Section A for details), which can create an anisotropic Floquet gap size in momentum space, as was already mentioned in the main text. Charge puddles could also play an important role in reducing the measured Hall response. The photocurrent $\Delta I$ flows through the most conducting puddles with larger chemical potential. Insulating regions with depleted photocurrent in this case would generate a much smaller Hall voltage than predicted, reducing the overall average Hall voltage.

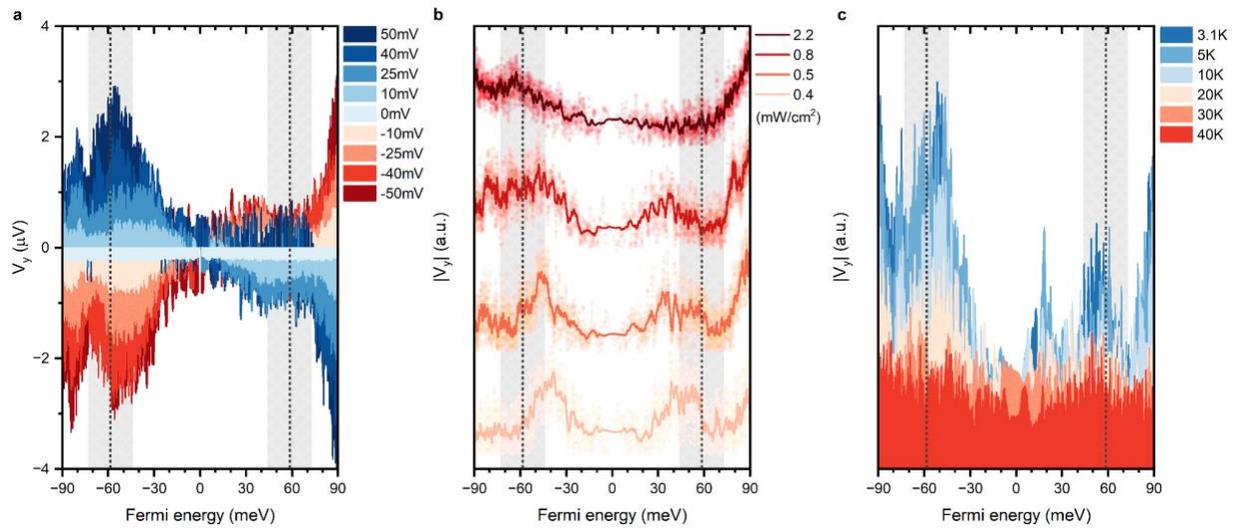

**Figure S10. Transverse voltage response under different conditions.** Transverse voltage as a function of $E_F$ under light irradiation ($\hbar\Omega = 117$ meV) with left-hand circular polarization. The dotted lines mark the values corresponding to $E_F = \pm 1/2\,\hbar\Omega$, with the related uncertainties indicated by the gray stripes. **a** Transverse voltage measured at different bias voltage values. The cryostat temperature is 3.1 K, and the laser beam is circularly polarized. **b** Transverse voltage measured under different irradiation power. The curves are vertically shifted for clarity. The cryostat temperature ranges from 3.0-3.3 K depending on laser intensity. The laser beam is circularly polarized. **c** Transverse voltage measured at different temperatures. For both **a** and **c**, the laser intensity is 0.8 mW/μm² and for **b** and **c** the bias voltage is 50 mV. The measurements are from sample C.

## L. Reproducibility of the results

The photoresponse features and the signatures of Floquet-Bloch band formation are reproducible with multiple sweeps on the same device. We also performed similar measurements on different devices. Figure **S11 a** and **b** show the Fermi energy dependence of the longitudinal photoresponse in samples B and C, respectively. Similar to sample A, the longitudinal photocurrent $\Delta I$ in both other samples also showed dips around $E_F = \pm 1/2\ \hbar\Omega$ for circularly and linearly polarized light. Such symmetric dips around the Dirac point are consistent with the emergent non-equilibrium Floquet steady states in graphene.

We also reproduced the transverse voltage as a function of Fermi energy in all our samples. Figure **S12** shows the transverse voltage as a function of $E_F$ under different polarizations of laser irradiation in sample C (panel **a**), and the difference in the Hall signal measured under opposite helicity of irradiation (panel **b**).

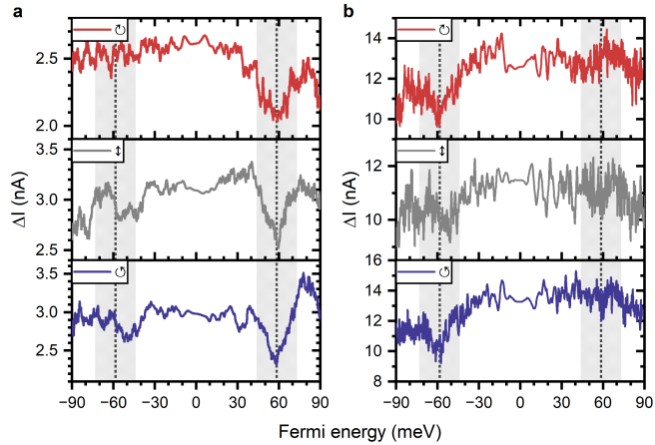

**Figure S11. Floquet signatures in the longitudinal photocurrent in additional devices.** The dotted lines mark the $E_F$ values corresponding to $\pm 1/2\ \hbar\Omega$, with the related uncertainty described by the gray stripes. **a** Longitudinal photocurrent as a function of $E_F$ under different polarization illumination in sample B. Parameters: Laser power density 1.4 mW/μm² and bias voltage 6 mV. **b** Longitudinal photocurrent as a function of $E_F$ under different polarization illumination in sample C. Parameters: Laser power density 1.6 mW/μm², bias voltage 6 mV, and cryostat temperature 3.4 K.

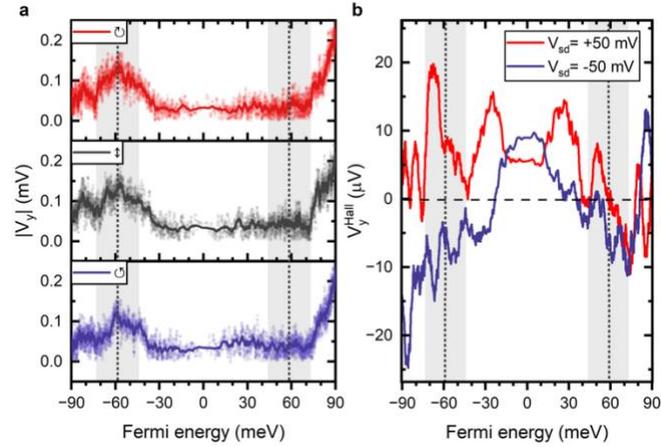

**Figure S12. Transverse voltage response in additional devices.** Transverse voltage as a function of $E_F$ under irradiation with different polarizations in sample C (panel **a**) and the difference between the transverse voltage response from circular polarizations of the laser beam with opposite chirality extracted from the curves in panel **a** (see panel **b**), and upon reversing the source-drain bias $V_{sd}$. Laser power density 0.5 mW/µm², cryostat temperature 3.1 K.

## M. Effect of Hall contact misalignment on the transverse signal measured under illumination

We made some estimates to quantify the contribution to the transverse voltage due to the geometric effect of the contact misalignment and compared it to the transverse voltage $V_y$ measured under illumination. First, we measured the transverse voltage $V_{Misalignment}$ as a function of the source-drain voltage $V_{SD}$ and the source-drain current $I_{SD}$ in the zero magnetic field, with no laser illumination and at room temperature. The measured misalignment for sample A is plotted in Figure **S13 a**. From this measurement we extracted the quantity $R_{Misalignment} = V_{Misalignment}/I_{SD} \sim 99.1$ Ω. Figure **S13 b** shows the estimated contribution to the transverse voltage from the photoinduced current change $\Delta I$, $V_{Misalignment} = R_{Misalignment} \Delta I$, in the same graph with the measured transverse photoresponse. The comparison shows that this misalignment contribution is much smaller than the transverse signal measured under illumination. Similar misalignment contributions were determined in all the measured samples.

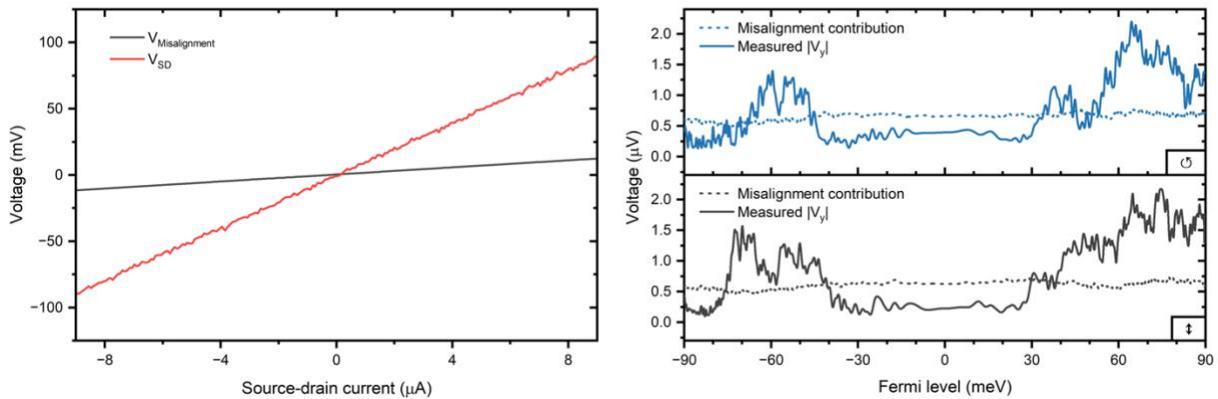

**Figure S13. Effect of Hall contacts misalignment. a** (Red) $V_{Misalignment}$ and (black) $V_{SD}$ measured at the same source-drain current. Room temperature. **b** Misalignment contribution to the measured transverse voltage for (top) circular and (bottom) linear irradiation. Solid lines represent the transverse voltages measured by the lock-in amplifier; dotted lines represent the contribution resulting from misalignment in the Hall contacts. The experimental data from Figure E1 of the main text are shown for reference.

## N. Linearity in the device response

The source-drain bias voltage dependence of the photoresponse is an important characteristic that can help determine its underlying physical mechanisms and separate intrinsic contributions from graphene or from the graphene-electrode boundaries. Figure **S14 a** shows a typical I-V curve (sample C). The clear linear relationship between the bias voltage and current indicates the ohmic contact nature of our graphene-electrode interface. Figure **S14 b** demonstrates the bias voltage dependence of the longitudinal photoresponse at a few different fixed $V_g$, which reveals a linear relationship between the sample response and the bias voltage. Moreover, no clear photoresponse signal was observed at $V_{sd} = 0$ for all measured gate voltages. Furthermore, as presented in Figure **S14 c**, the normalized photoresponse curves collected at different $V_{sd}$ as a function of Fermi energy further confirm the linearity of our sample's photoresponse.

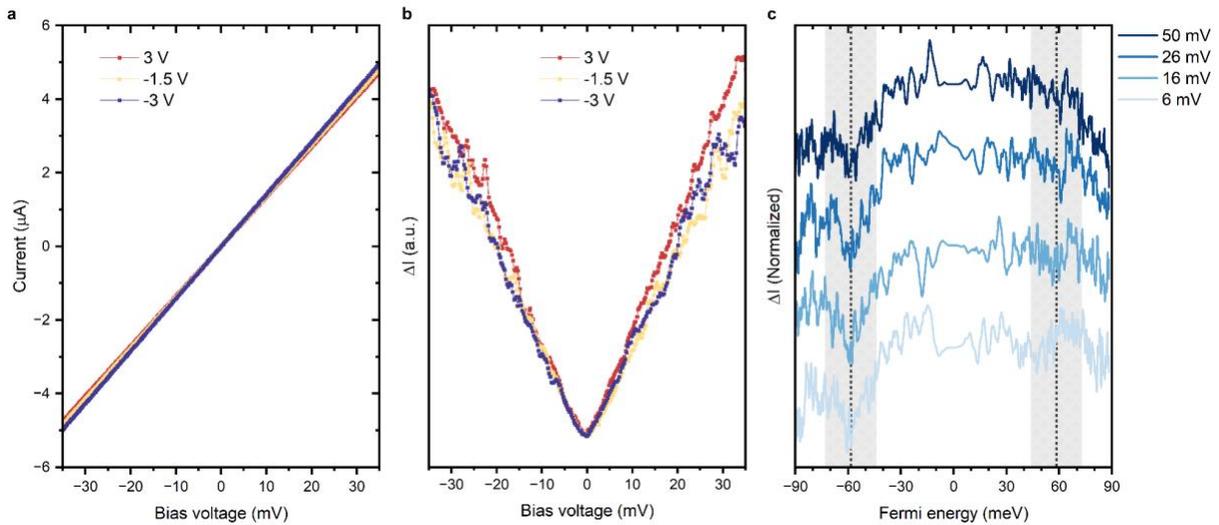

**Figure S14. Bias dependence of sample response. a** I-V curve from sample C at different values of $V_g$. **b** Photo-induced current change $\Delta I$ as a function of bias voltage at different values of $V_g$. **c** Normalized longitudinal photoresponse $\Delta I$ as a function of $E_F$ at different bias voltages, under circularly polarized irradiation, with a power density of 1.5 mW/µm². The curves are vertically shifted for clarity. The cryostat temperature is 3.4 K.

## O. Chopper frequency dependence of the photoresponse

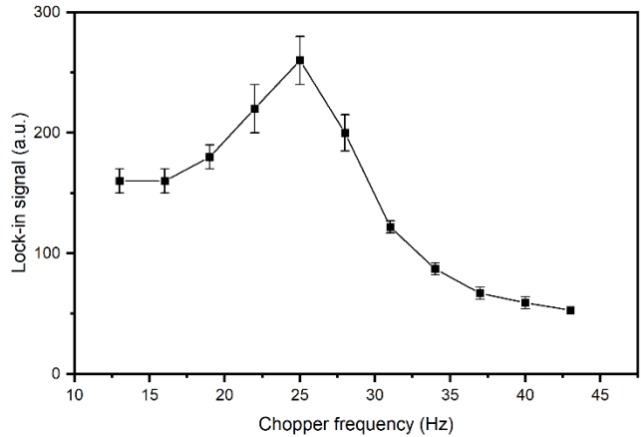

**Figure S15. Longitudinal photoresponse as a function of chopper frequency.** Photo-induced current change measured in sample C under 1.5 mW/μm² of circularly polarized irradiation. The sample is biased at a fixed 6 mV. The cryostat temperature is 3.4 K.

To better understand our samples' photoresponse, we measured the longitudinal photoresponse as a function of the chopper modulation frequency. As one can see in Figure **S15**, as the chopper frequency is increased, the signal first increases and then rapidly drops. Such non-monotonic behavior can be attributed to the heating and cooling processes of electrons that are associated with the chopping of the laser beam. At low frequencies, electrons are heated with a longer exposure time to the irradiation during each chopping cycle, while for high frequencies, although the exposure time is shorter, the electrons do not have enough time for cooling. As a result, for both too low and too high frequencies, a higher effective electron temperature is expected. Furthermore, as discussed in the main text, an elevated effective electron temperature will not only lead to the fading of Floquet features but will also suppress the overall signal strength. Therefore, for better detection of the Floquet features, one has to carefully control the chopper frequency and maintain the electron temperatures below the size of Floquet gaps. All results presented in this work were collected with a chopper frequency around 23 Hz.

## P. Error Analysis

In this section, we provide details of the error estimates for sample C, as presented in Figure 3 **c-d** in the main text. The electronic carrier density in the sample was estimated using the relation

$$n = k(V_g - V_d),$$

where $k \approx 1.63 \times 10^{-15}$ m$^{-2}$/V $\pm \delta k$ and $\delta k = 9 \times 10^{13}$ m$^{-2}$/V were estimated from the gate efficiency calibration. The voltage $V_d \approx 0.55\ V \pm \delta V_d$ was estimated to be the gate voltage at which $\sigma_{xx}$ is minimized, and the uncertainty $\delta V_d \approx 0.8\ V$ was estimated from the gate efficiency calibration. The corresponding uncertainties in the Fermi energy $E_F = \hbar v_F \sqrt{\pi n}$ are given by $\delta E_F \approx (E_F/2)\sqrt{(\delta k/k)^2 + [\delta V_d/(V_g - V_d)]^2}$.

The full width at half maximum (FWHM) of the conductivity dip (see Figure 3 **d**) $FWHM = \Delta E_F = E_F(V_{g1}) - E_F(V_{g2})$, where $V_{g1} = B + \sqrt{2 \ln 2}\ \sigma$ and $V_{g2} = B - \sqrt{2 \ln 2}\ \sigma$, was estimated by performing a least squares fit of the experimental data to the function

$$\Delta I_{fit} = a_0 \exp(-(V_g - B)^2/2\sigma^2) + CV_g + D.$$

We estimated the uncertainty in the FWHM using the relationship

$$\delta \Delta E_F = \sqrt{\left(\frac{\partial \Delta E_F}{\partial v_F}\delta v_F\right)^2 + \left(\frac{\partial \Delta E_F}{\partial V_d}\delta V_d\right)^2 + \left(\frac{\partial \Delta E_F}{\partial k}\delta k\right)^2 + \left(\frac{\partial \Delta E_F}{\partial V_{g1}}\delta V_{g1}\right)^2 + \left(\frac{\partial \Delta E_F}{\partial V_{g2}}\delta V_{g2}\right)^2},$$

where $\delta v_F \approx 0.008 \times 10^6$ m/s as determined by cyclotron resonance measurements. The uncertainties $\delta V_{g1}$ and $\delta V_{g2}$ were estimated using the relation

$$\delta V_{g1} \approx \delta V_{g2} \approx \sqrt{\delta B^2 + (\sqrt{2 \ln 2}\ \delta\sigma)^2}$$

where $\delta B$ and $\delta\sigma$ are the uncertainties in the fitting parameters $B$ and $\sigma$, respectively, as determined from the 68% confidence interval of the least squares fit.

## References

1  KOMINAMI, M., POZAR, D. & SCHAUBERT, D. DIPOLE AND SLOT ELEMENTS AND ARRAYS ON SEMI-INFINITE SUBSTRATES. *IEEE TRANSACTIONS ON ANTENNAS AND PROPAGATION* **33**, 600-607 (1985).
2  Biagioni, P., Huang, J., Duò, L., Finazzi, M. & Hecht, B. Cross Resonant Optical Antenna. *PHYSICAL REVIEW LETTERS* **102** (2009). https://doi.org/10.1103/PhysRevLett.102.256801
3  Bolotin, K. I., Sikes, K. J., Hone, J., Stormer, H. L. & Kim, P. Temperature-dependent transport in suspended graphene. *Physical Review Letters* **101** (2008). https://doi.org/10.1103/PhysRevLett.101.096802

4       Chen, J. H. *et al.* Charged-impurity scattering in graphene. *Nature Physics* **4**, 377-381 (2008). https://doi.org/10.1038/nphys935
5       Somphonsane, R. *et al.* Evaluating the Sources of Graphene's Resistivity Using Differential Conductance. *Scientific Reports* **7** (2017). https://doi.org/10.1038/s41598-017-10367-1
6       Dean, C. *et al.* Boron nitride substrates for high-quality graphene electronics. *NATURE NANOTECHNOLOGY* **5**, 722-726 (2010). https://doi.org/10.1038/nnano.2010.172
7       Castro Neto, A. H., Guinea, F., Peres, N. M. R., Novoselov, K. S. & Geim, A. K. The electronic properties of graphene. *Reviews of Modern Physics* **81**, 109-162 (2009). https://doi.org/10.1103/RevModPhys.81.109
8       Guinea, F., Katsnelson, M. I. & Vozmediano, M. A. H. Midgap states and charge inhomogeneities in corrugated graphene. *Physical Review B* **77** (2008). https://doi.org/10.1103/PhysRevB.77.075422
9       Martin, J. *et al.* Observation of electron-hole puddles in graphene using a scanning single-electron transistor. *Nature Physics* **4**, 144-148 (2008). https://doi.org/10.1038/nphys781
10      McCann, E. *et al.* Weak-localization magnetoresistance and valley symmetry in graphene. *Physical Review Letters* **97** (2006). https://doi.org/10.1103/PhysRevLett.97.146805
11      Wu, X. S., Li, X. B., Song, Z. M., Berger, C. & de Heer, W. A. Weak antilocalization in epitaxial graphene: Evidence for chiral electrons. *Physical Review Letters* **98** (2007). https://doi.org/10.1103/PhysRevLett.98.136801
12      Cleary, J. W., Smith, E. M., Leedy, K. D., Grzybowski, G. & Guo, J. P. Optical and electrical properties of ultra-thin indium tin oxide nanofilms on silicon for infrared photonics. *Optical Materials Express* **8**, 1231-1245 (2018). https://doi.org/10.1364/ome.8.001231
13      Genty, F. *et al.* in *MRS.*  **1327** (*MRS Online Proceedings Library*).
14      Chuang, K. C., Deacon, R. S., Nicholas, R. J., Novoselov, K. S. & Geim, A. K. Cyclotron resonance of electrons and holes in graphene monolayers. *Philosophical Transactions of the Royal Society a-Mathematical Physical and Engineering Sciences* **366**, 237-243 (2008). https://doi.org/10.1098/rsta.2007.2158
15      Russell, B. J., Zhou, B. Y., Taniguchi, T., Watanabe, K. & Henriksen, E. A. Many-Particle Effects in the Cyclotron Resonance of Encapsulated Monolayer Graphene. *Physical Review Letters* **120** (2018). https://doi.org/10.1103/PhysRevLett.120.047401
16      Abanin, D. A. & Levitov, L. S. Conformal invariance and shape-dependent conductance of graphene samples. *Physical Review B* **78** (2008). https://doi.org/10.1103/PhysRevB.78.035416
17      Yang, Y. F. *et al.* Low Carrier Density Epitaxial Graphene Devices On SiC. *Small* **11**, 90-95 (2015). https://doi.org/10.1002/smll.201400989
18      El Fatimy, A. *et al.* Epitaxial graphene quantum dots for high-performance terahertz bolometers. *Nature Nanotechnology* **11**, 335-+ (2016). https://doi.org/10.1038/nnano.2015.303
19      Mhatre, S. M. *et al.* Dynamics of transient hole doping in epitaxial graphene. *Physical Review B* **105** (2022). https://doi.org/10.1103/PhysRevB.105.205423
20      Rudner, M. S. & Lindner, N. H. in *arXiv:2003.08252v2 [cond-mat.mes-hall]*   (2020).
21      Oka, T. & Aoki, H. Photovoltaic Hall effect in graphene (vol 79, 081406, 2009). *Physical Review B* **79** (2009). https://doi.org/10.1103/PhysRevB.79.169901

22    Kitagawa, T., Oka, T., Brataas, A., Fu, L. & Demler, E. Transport properties of nonequilibrium systems under the application of light: Photoinduced quantum Hall insulators without Landau levels. *Physical Review B* **84** (2011). https://doi.org/10.1103/PhysRevB.84.235108
23    Calvo, H. L., Pastawski, H. M., Roche, S. & Torres, L. Tuning laser-induced band gaps in graphene. *Applied Physics Letters* **98** (2011). https://doi.org/10.1063/1.3597412
24    Seetharam, K. I., Bardyn, C. E., Lindner, N. H., Rudner, M. S. & Refael, G. Controlled Population of Floquet-Bloch States via Coupling to Bose and Fermi Baths. *Physical Review X* **5** (2015). https://doi.org/10.1103/PhysRevX.5.041050
25    Seetharam, K., Bardyn, C., Lindner, N., Rudner, M. & Refael, G. Steady states of interacting Floquet insulators. *PHYSICAL REVIEW B* **99** (2019). https://doi.org/10.1103/PhysRevB.99.014307
26    Esin, I., Rudner, M., Refael, G. & Lindner, N. Quantized transport and steady states of Floquet topological insulators. *PHYSICAL REVIEW B* **97** (2018). https://doi.org/10.1103/PhysRevB.97.245401
27    Yang, C., Esin, I., Lewandowski, C. & Refael, G. Optical Control of Slow Topological Electrons in Moire Systems. *PHYSICAL REVIEW LETTERS* **131** (2023). https://doi.org/10.1103/PhysRevLett.131.026901
28    Li, Q., Hwang, E. & Das Sarma, S. Disorder-induced temperature-dependent transport in graphene: Puddles, impurities, activation, and diffusion. *PHYSICAL REVIEW B* **84** (2011). https://doi.org/10.1103/PhysRevB.84.115442
29    McIver, J. W. *et al.* Light-induced anomalous Hall effect in graphene. *Nature Physics* **16**, 38-+ (2020). https://doi.org/10.1038/s41567-019-0698-y
30    Candussio, S. *et al.* Terahertz radiation induced circular Hall effect in graphene. *Physical Review B* **105** (2022). https://doi.org/10.1103/PhysRevB.105.155416
31    Durnev, M. V. Photovoltaic Hall effect in the two-dimensional electron gas: Kinetic theory. *Physical Review B* **104** (2021). https://doi.org/10.1103/PhysRevB.104.085306